\documentclass[prb,twocolumn,showpacs,preprintnumbers,float]{revtex4-1}

\usepackage[colorlinks=true,urlcolor=blue,citecolor=blue,linkcolor=blue,breaklinks=true]{hyperref}
\usepackage{graphicx}
\usepackage{amsmath}
\usepackage{amssymb}
\usepackage{times}
\usepackage{color}
\usepackage{ulem}
\usepackage{physics}
\usepackage{bm}
\usepackage{cleveref}
\usepackage{xr}
\usepackage{multirow}
\usepackage{comment}

\newcommand{\hr}[1]{{\color{black} #1}}

\definecolor{caribbeangreen}{rgb}{0.0, 0.8, 0.6}

\usepackage[utf8]{inputenc}
\usepackage{siunitx}

\begin{document}

\title{Quantum and temperature effects on crystal structure of superhydride: A path integral molecular dynamics study}%

\author{Yuta Watanabe$^1$}%
\email{watanabe-yuta613@g.ecc.u-tokyo.ac.jp}%
\author{Takuya Nomoto$^2$}%
\email{nomoto@ap.t.u-tokyo.ac.jp}%
\author{Ryotaro Arita$^{2,3}$}%
\affiliation{$^1$Department of Applied Physics, The University of Tokyo, Hongo, Bunkyo-ku, Tokyo, 113-8656, Japan \\
$^2$ Research Center for Advanced Science and Technology, University of Tokyo, Komaba Meguro-ku, Tokyo 153-8904, Japan\\
$^3$RIKEN Center for Emergent Matter Science (CEMS), Wako 351-0198, Japan}
\date{\today}%

\begin{abstract}
By classical and path-integral molecular dynamics simulations, we study the pressure-temperature ($P$-$T$) phase diagram of LaH$_{10}$ to clarify the impact of temperature and atomic zero-point motions. We calculate the XRD pattern and analyze the space group of the crystal structures. For 125 GPa $\leq P\leq$ 150 GPa and $T=300$ K, we show that a highly symmetric $Fm\bar{3}m$ structure, for which superconductivity is particularly favored, is stabilized only by the temperature effect. On the other hand, for $T=200$ K, the interplay between the temperature and quantum effects is crucial to realize the $Fm\bar{3}m$ structure. For $P=$100 GPa and $T=$300 K, we find that the system is close to the critical point of the structural phase transition between the $Fm\bar{3}m$ structure and those with lower symmetries.
\end{abstract}

\maketitle

\section{Introduction}



In his seminal paper of 1968, Ashcroft proposed that metallic hydrogen under extremely high pressure will exhibit superconductivity at  very high temperatures due to its strong electron-phonon coupling and high phonon frequencies~\cite{ashcroft_metallic_1968.org}. However, the experimental realization of metallic hydrogen turned out to be a formidable challenge, and it has been a holy grail in the field of high-pressure experiments~\cite{monacelli_black_2021,eremets_conductive_2011,azadi_dissociation_2014,cudazzo_electron-phonon_2010,cudazzo_electron-phonon_2010-1,dalladay-simpson_evidence_2016,mcminis_molecular_2015,dias_observation_2017,Loubeyre_2020,mcmahon_properties_2012}. 
On the other hand, when hydrogen atoms are embedded in a network of other elements, 
the chemical precompression can make a situation similar to that of metallic hydrogen with much lower pressure~\cite{gilman_lithium_1971,ashcroft_hydrogen_2004}. Inspired by this idea,
a variety of hydrides have been investigated experimentally and theoretically~\cite{flores-livas_perspective_2020}. 
In particular, numerical simulations have made crucial contributions in searching for thermodynamically and dynamically stable materials and predicting their superconducting transition temperatures ($T_c$'s)~\cite{quan_compressed_2019,feng_compressed_2015,ge_first-principles_2016,errea_high-pressure_2015,zurek_high-temperature_2019,papaconstantopoulos_high-temperature_2020,peng_hydrogen_2017,di_cataldo_bh_2021,liu_microscopic_2019,gao_phonon-mediated_2021,liu_potential_2017,shipley_stability_2020,kim_predicted_2009,verma_prediction_2021,li_pressure-stabilized_2015,sun_route_2019,yi_stability_2021,kruglov_superconductivity_2020}.

While there have been many reports on the discovery of high $T_{c}$ hydride superconductors so far~\cite{duan_pressure-induced_2015,troyan_anomalous_2021,drozdov_conventional_2015,einaga_crystal_2016,ma_high-tc_2021,chen_high-temperature_2021,snider_room-temperature_2020,semenok_superconductivity_2020,semenok_superconductivity_2021,matsuoka_superconductivity_2019,somayazulu_evidence_2019,drozdov_superconductivity_2019},
$\mathrm{LaH_{10}}$ is a prototypical hydrogen-based superconductor for which 
$T_{c}\sim$ \SI{250}{K} has been observed~\cite{somayazulu_evidence_2019,drozdov_superconductivity_2019,sun_high-temperature_2021}. Interestingly, its $T_c$ sensitively depends on the crystal structure, 
and it has been shown that the $Fm\bar{3}m$ structure is exceptionally favorable for phonon-mediated superconductivity.
However, this $Fm\bar{3}m$ structure is not robust under low pressure and easily distorts to low-symmetry structures such as $C2/m$ or $R\bar{3}m$, for which high $T_c$ cannot be expected~\cite{geballe_synthesis_2018, liu_potential_2017, errea_quantum_2020,sun_high-temperature_2021}. 
In the previous first-principles calculations based on density functional theory (DFT), the critical pressure has been estimated to be higher than $\sim$\SI{200}{GPa}.
On the other hand, it has been experimentally suggested that the $Fm\bar{3}m$ structure survives down to $\sim$ \SI{150}{GPa}~\cite{drozdov_superconductivity_2019, somayazulu_evidence_2019}.

For this issue, there are two effects that should be considered seriously but are neglected in standard calculations based on density functional theory (DFT). One is the effect of quantum fluctuations of hydrogen nuclei, and the other is that of finite temperature. While the former has been shown to stabilize the $Fm\bar{3}m$ structure~\cite{errea_quantum_2020}, the latter is yet to be fully understood.

For the effect of finite temperature, hydrogen diffusion has been investigated in several studies. While hydrogen atoms can diffuse in the clathrate superhydride $\mathrm{Li_{2}MgH_{16}}$ even at low temperature~\cite{wang_quantum_2021}, protons in $\mathrm{LaH_{10}}$ get diffusive only at temperatures higher than \SI{800}{K} at \SI{150}{GPa}~\cite{liu_dynamics_2018}. 
However, this does not necessarily indicate that the temperature effect plays an irrelevant role in determining the symmetry of the crystal structure. Indeed, the temperature is one of the critical parameters for synthesis processes in the experiments, and accurate determination of the pressure-temperature ($P$-$T$) phase diagram is highly desired.

To elucidate the finite-$T$ effect on the crystal structure, in the present study, we perform a molecular dynamics (MD) calculation for $\mathrm{LaH_{10}}$ and analyze the symmetry of the crystal structure at thermodynamic equilibrium. When the quantum effect of hydrogen nuclei is neglected, we show that the $Fm\bar{3}m$ structure is stable even at $P$=125$\sim$150 GPa if the temperature is higher than \SI{300}{K}. We then take account of the quantum effect by a calculation based on the path integral molecular dynamics (PIMD)\footnote{For MD and PIMD simulations to discuss the dynamical properties of $\mathrm{LaH_{10}}$, see Ref.~\onlinecite{liu_dynamics_2018}}. We confirm that the quantum effect expands the $Fm\bar{3}m$ phase further in the $P$-$T$ phase diagram. Significantly, the $Fm\bar{3}m$ structure becomes stable at $P$=125$\sim$150 GPa and $T$=200 K, consistent with the experimental observation. We also calculate the X-ray diffraction (XRD) spectrum and find that the distortion at $P$=100 GPa becomes weak in the PIMD calculation. These results indicate that not only the quantum effect but the finite-$T$ effect also plays a crucial role in forming the $Fm\bar{3}m$ structure.

\section{method}
In order to obtain a thermodynamically stable crystal structures considering the quantum nuclear effect at finite temperature, we performed PIMD simulations. We combined the PIMD and DFT, which is called $ab$ $initio$ PIMD. In this section, we review the scheme of PIMD and explain the simulation details.

Let us consider a $N$-body Hamiltonian with a potential $V$ given as
\begin{equation}
    H = \sum_{i}^{N}\frac{{\bm p}_{i}^2}{2m_{i}} + V({\bm r}_{1}, \cdots, {\bm r}_{N}),
\end{equation}
where {${\bm p}_{i}$}, $m_{i}$ and ${\bm r}_{i}$ are the momentum, mass, and position of the $i$-th particle, respectively. The partition function of this system can be written as
\begin{equation}
    Z = \mathrm{Tr}\exp(-\beta H),
\end{equation}
where $\beta = 1/k_{B} T$ and {$k_{B}$} is the Boltzmann constant. By using the Suzuki-Trotter decomposition, this partition function can be written as
\begin{multline}
    Z = \lim_{P_b\rightarrow \infty}\Pi_{i = 1}^{N}\left[\left(\frac{m_{i}P_b}{2\pi \beta \hbar^{2}}\right)^{3P_b/2}\Pi_{j = 1}^{P_b}\int d\bm{r}_{i,j}\right. \\
    \exp \left[-\beta \sum_{j = 1}^{P_b}\left[\sum_{i = 1}^{N}\frac{m_{i}P_b}{2(\beta\hbar)^{2}}(\bm{r}_{i,j} - \bm{r}_{i,j-1})^{2} \right. \right.\\
    +\left. \left. \left. \frac{1}{P_b}V(\bm{r}_{1,j}, \cdots, \bm{r}_{N,j})\right]\right]\right].
\label{eq:partitionfun}
\end{multline}
We see that the original $N$ body quantum partition function can be regarded as a $NP_b$ body classical partition function where the potential is given as
\begin{multline}
    V_{\mathrm{PI}} = \sum_{j = 1}^{P_b}\left[\sum_{i = 1}^{N}\frac{m_{i}P_b}{2(\beta\hbar)^{2}}(\bm{r}_{i,j} - \bm{r}_{i,j-1})^{2} \right. \\
    + \left. \frac{1}{P_b}V(\bm{r}_{1,j}, \cdots, \bm{r}_{N,j})\right].
\end{multline}

\begin{figure}[htbp]
 \center{
 \includegraphics[keepaspectratio, scale=0.6]{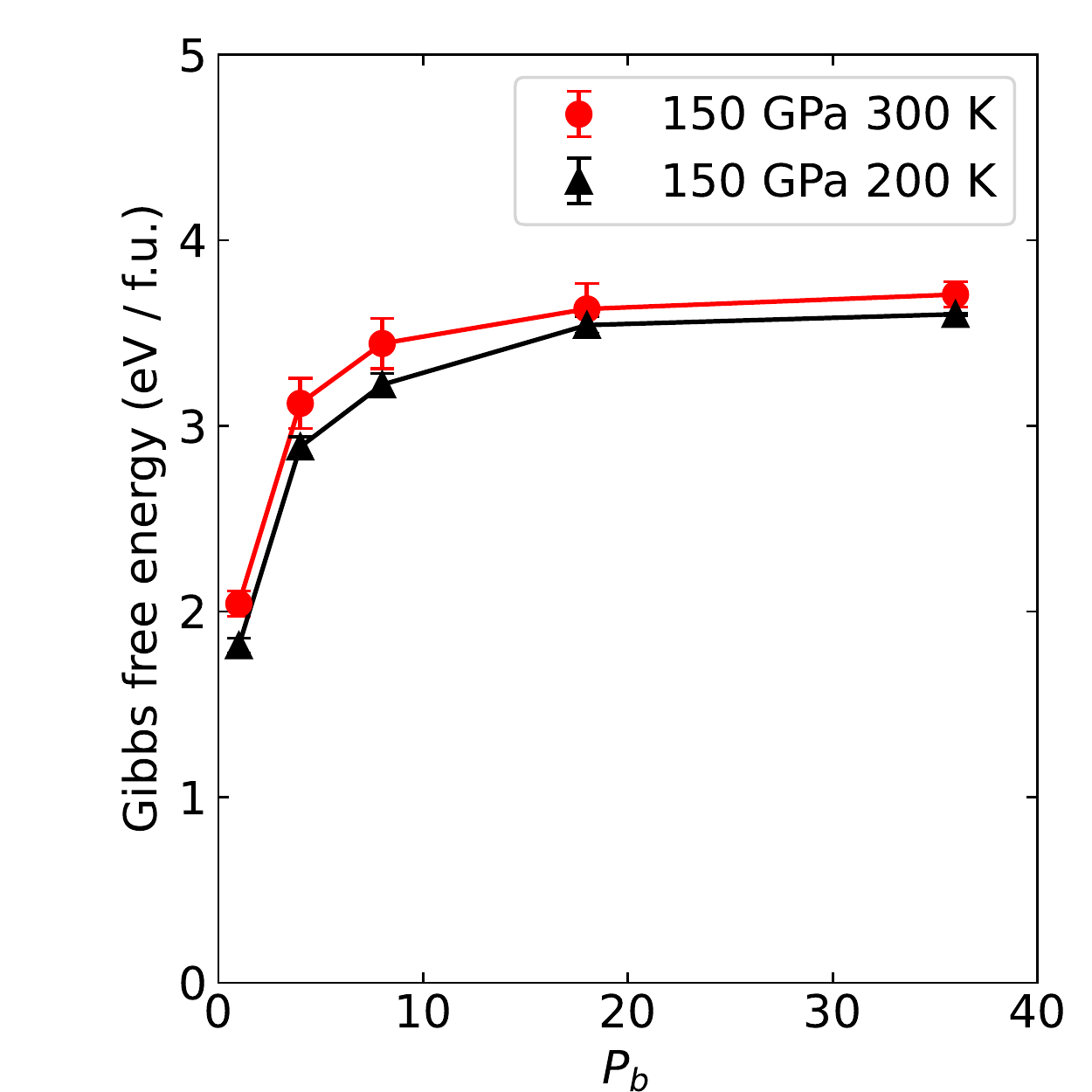}}
 \caption{$P_b$ (the number of the Suzuki-Trotter decomposition in eq.~\ref{eq:partitionfun}) dependence of the Gibbs free energy at $(P, T) =$ (\SI{150}{GPa}, \SI{300}{K}) \hr{and (\SI{150}{GPa}, \SI{200}{K}). The last 250 fs of the simulation are divided into ten bins for 200 K. Then, the average and standard deviation of the sum of the energy and the $PV$ term for each bin are calculated. For 300 K, we took 750 fs for the MD calculation to reduce the error.} Regarding the entropy term, we evaluated it from the velocity of each atom~\cite{entropy}. We used a $4\times 4\times 4$ $\bm{k}$-mesh in the PIMD simulation here. \label{ent}}
 
\end{figure}

We performed PIMD simulation~\cite{shiga_unified_2001,tuckerman,marx} on these $NP_b$ particles in a system containing four formula units of $\mathrm{LaH_{10}}$ with $N = 44$ atoms and $P_b = 36$ using the PIMD software package~\cite{pimd} and Vienna $ab$  $initio$ simulation package (VASP)~\cite{kresse_efficient_1996} based on the projector-augmented wave (PAW) method~\cite{kresse_ultrasoft_1999,blochl_projector_1994}. We used the exchange correlation functional in the the generalized-gradient approximation (GGA) proposed by Perdew, Burke and Ernzerhof (PBE)~\cite{perdew_generalized_1996}
\footnote{For the dependence of the critical pressure at the structural phase transition on the choice of the pseudopotentials and exchange-correlation functionals, see Ref.~\onlinecite{sura_effects_2022}}. The cutoff energy was set to be \SI{700}{eV}, with a $4\times 4\times 4$ and $6\times 6\times 6$ $\bm{k}$-mesh. Hereafter, we mainly show the results for the $6\times 6\times 6$ $\bm{k}$-mesh. The typical time-steps $\Delta t$ was set to 0.25 $\sim$ \SI{0.5}{fs} in the MD calculation. 
In Fig.~\ref{ent}, we plot the $P_b$ dependence of the Gibbs free energy of the system. We see that the calculation well converges \hr{at both 200 K and 300 K}, and the quantum effect (the zero-point energy of the atoms) is accurately considered for $P_b=36$. \hr{The convergence with the number of atoms is also checked by examining the crystal structures obtained from classical molecular dynamics simulations with eight formula units of $\mathrm{LaH_{10}}$ for ($P, T$) = (150 GPa, 300 K) and (125 GPa, 200 K). We found that the phase boundary between the symmetric and distorted structures does not change, indicating that the supercell of 44 atoms is large enough to discuss the phase diagram.}

\begin{figure}[htbp]
 \center{
 \includegraphics[keepaspectratio, scale=0.6]{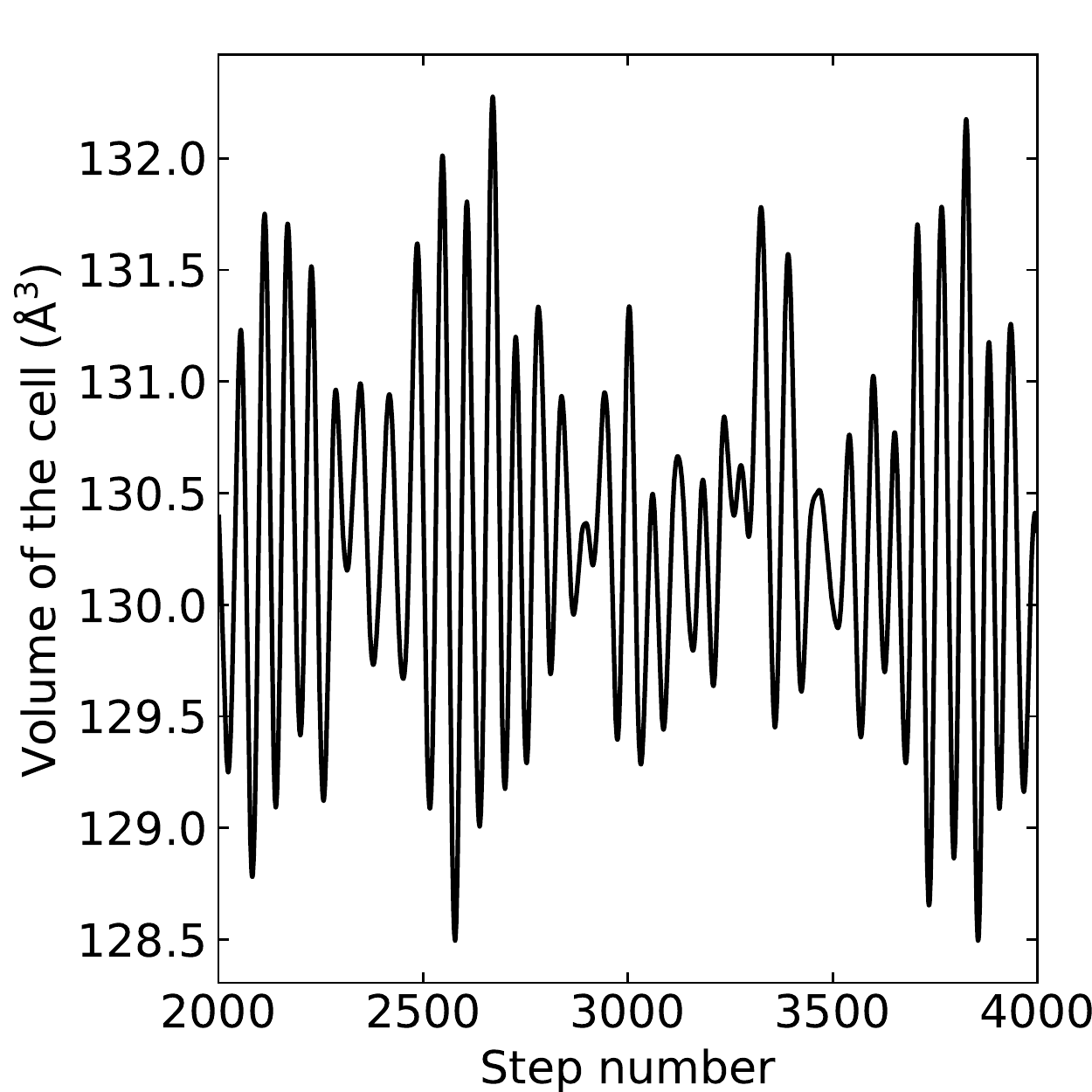}}
 \caption{Time evolution of the volume of the $\mathrm{LaH_{10}}$ crystal for $(P, T) = $ (\SI{150}{GPa}, \SI{300}{K}) in the classical MD simulation.\label{vol}}
 
\end{figure}

In the MD calculation, we move the ions following the electronic potential in the real space. At each step, the resulting crystal structure fluctuates around the ideal crystal structure in thermodynamic equilibrium. 
In Fig.~\ref{vol}, we plot the time evolution of the volume of the system. We see that the amplitude of the deviation from the equilibrium value is about $\sim$2\AA$^3$.

To reduce the error due to this fluctuation, 
we adopted the following method. When we obtain a crystal structure at the $n_0$-th step, 
we calculate $F(n_{0},n_{0} + n_{\rm avg})$, which is a crystal structure averaged from the $n_{0}$-th to $(n_{0} + n_{\rm avg})$-th step. For $n_{\rm avg}$, we took a thousand to several ten thousand steps. We then compared the X-ray diffraction (XRD) patterns of $F(n_{0},n_{0} + n_{\rm avg})$ with several different $n_{0}$ and $n_{\rm avg}$. The crystal structure is in thermodynamic equilibrium when only minor differences in XRD patterns are observed. \hr{We did not observe the hydrogen diffusion discussed in Ref.~\onlinecite{liu_dynamics_2018}, 
because the temperatures in our calculations are much lower than that in the previous study where the hydrogen diffusion occurs.}

On top of the XRD simulation, we determine the space group of the obtained averaged crystal structures to discuss the symmetry of the crystals. We used VESTA~\cite{vesta} for XRD simulation and pymatgen module of pymatgen.symmetry.analyzer.SpacegroupAnalyzer for the space group analysis~\cite{ong_python_2013,togo_textttspglib_2018}.

\section{Results and Discussion}

\begin{figure}[htbp]
 \center{
 \includegraphics[keepaspectratio, scale=0.3]{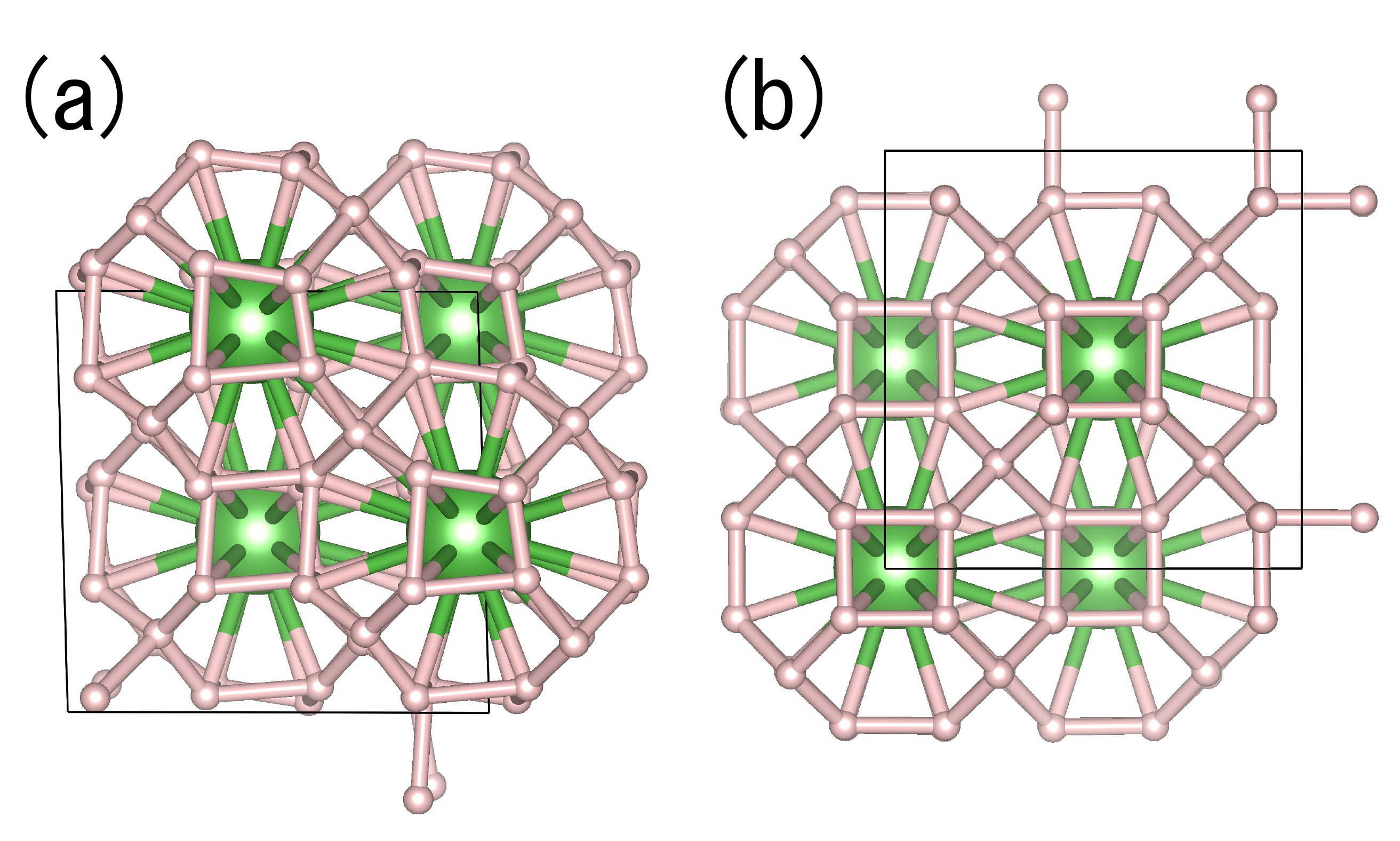}}
 \caption{Crystal structure of $\mathrm{LaH_{10}}$ obtained from the classical MD simulation for ($P$,$T$)=(150 GPa, 200 K) (a) and that for (150 GPa, 300 K) belonging to the space group of $Fm\bar{3}m$ (b). 
 \label{crystal}}
 
\end{figure}

Let us first look at the calculated crystal structure of $\mathrm{LaH_{10}}$ \hr{obtained from classical MD simulation}. In Fig.~\ref{crystal}, we compare the structure for ($P$,$T$)=(150 GPa, 200 K) and that for (150 GPa, 300 K). We see that the former is severely distorted, while the latter has a highly symmetric ($Fm\bar{3}m$) structure.


\begin{figure}[htbp]
        \centering
        \includegraphics[keepaspectratio, scale=0.65]{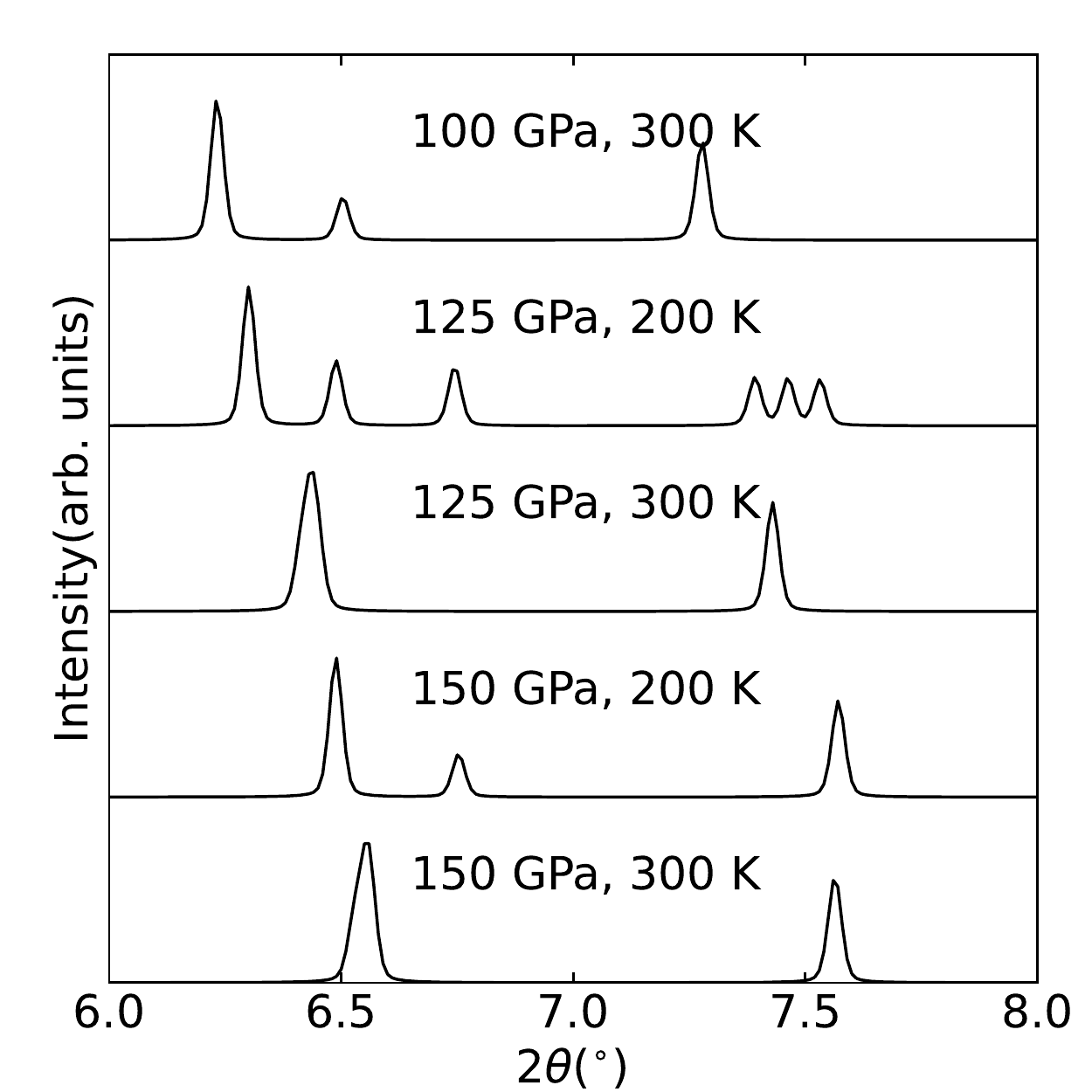}
        
        \caption{XRD patterns of $\mathrm{LaH_{10}}$ crystals obtained by \hr{classical} MD around $2 \theta = 6.5^{\circ}$ with the wavelength of $\lambda = 0.3344$ \AA. \label{md_xrd}}
\end{figure}

\begin{figure}
        \centering
        \includegraphics[keepaspectratio, scale=0.65]{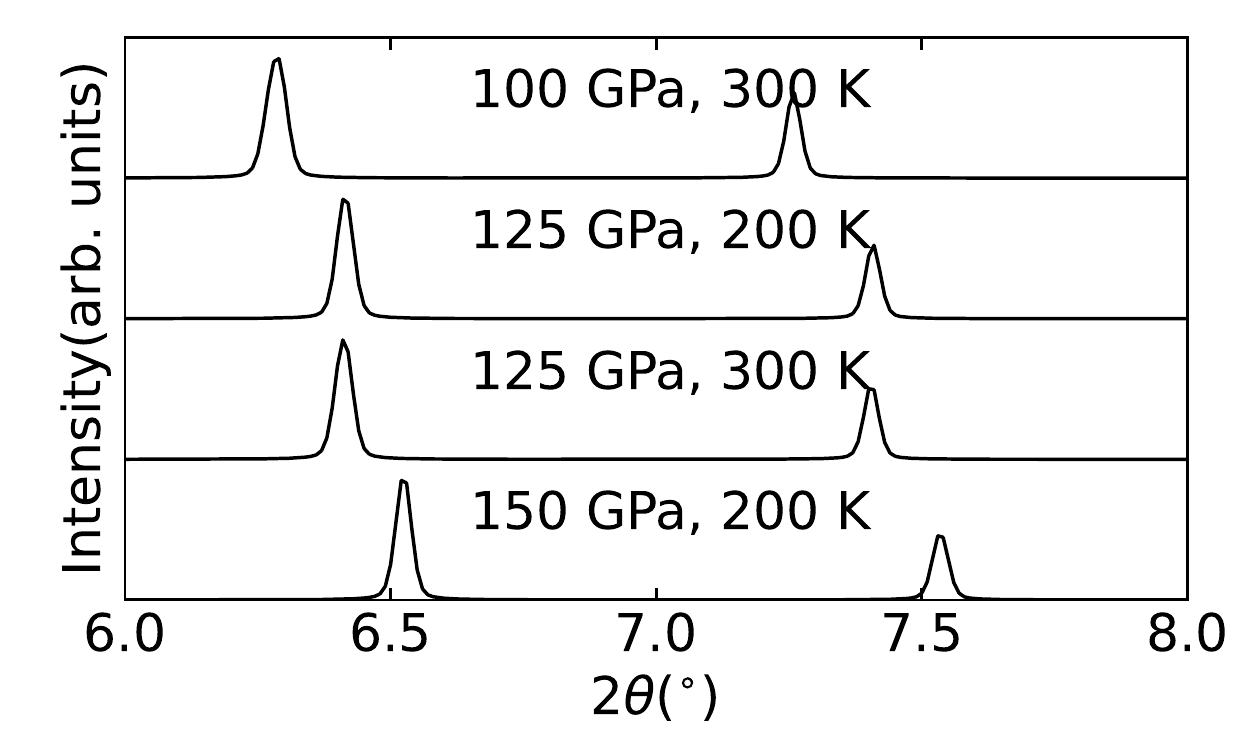}
        
    \caption{XRD patterns of $\mathrm{LaH_{10}}$ crystals obtained by PIMD around $2 \theta = 6.5^{\circ}$ with the wavelength of $\lambda = 0.3344$ \AA. \label{pimd_xrd}}
\end{figure}

 In Figs.~\ref{md_xrd} and \ref{pimd_xrd}, we show the X-ray diffraction (XRD) patterns for the time-step-averaged crystal structure calculated by the classical and path-integral MD simulation, respectively. In Ref.~\onlinecite{errea_quantum_2020}, it has been shown that
 the distorted low-symmetry structure with the space group of $C2$ and $R\bar{3}m$ show a characteristic split of the peak around $2 \theta = 6.5^{\circ}$ for the wavelength $\lambda =$ \SI{0.3344}{\AA}. Hereafter we focus on this peak splitting in this region as an indicator of whether the crystal structure is distorted or not.

In Fig.~\ref{md_xrd}, we see that the split of the peaks around $2 \theta = 6.5^{\circ}$ exists for $(P, T) = $ (\SI{100}{GPa}, \SI{300}{K}), (\SI{125}{GPa}, \SI{200}{K}), and  (\SI{150}{GPa}, \SI{200}{K}). On the other hand, this split disappears for $(P, T) = $ (\SI{125}{GPa}, \SI{300}{K}) and (\SI{150}{GPa}, \SI{300}{K}). Namely, at $P=$\SI{125} and \SI{150}{GPa}, while the crystal structure is severely distorted at $T=$\SI{200}{K}, it becomes symmetric at $T=$\SI{300}{K}. This temperature effect is not effective for $P=$\SI{100}{GPa} even at $T=$\SI{300}{K}.
As for the monotonic shift of the signal along with the pressure change, it can be understood in terms of the systematic compression of the crystal. 

In Fig.~\ref{pimd_xrd}, we show how the synergy effect of the zero-point motion of atoms and finite temperature can be seen in the PIMD simulation. In contrast with Fig.~\ref{md_xrd}, we see that there is no split for $(P, T) = $ (\SI{100}{GPa}, \SI{300}{K}), (\SI{125}{GPa}, \SI{200}{K}), and  (\SI{150}{GPa}, \SI{200}{K}). This result indicates that the symmetric $Fm\bar{3}m$ structure survives even at \SI{100}{GPa}, if the temperature is as high as \SI{300}{K}. \hr{It should be noted that the MD results always contain stochastic errors. Therefore, we cannot determine the space group by just looking at the XRD data. Thus, we combine the XRD analysis and space group analysis in the present study.}


Let us now move on to the space group analysis using the pymatgen module. In Table~\ref{pymatgen}, we list the results for the calculated structures in Figs.~\ref{md_xrd} and \ref{pimd_xrd}. Here, "symprec" 
is a threshold used to identify the space group. Smaller symprec means that the space group is determined more strictly. If the crystal structure is judged to have a high symmetry only for large symprec, the crystal structure should have some distortion. For example, the space group of the crystal structure obtained by PIMD for $(P, T) = $ (\SI{100}{GPa}, \SI{300}{K}) is $Fm\bar{3}m$ only for symprec $\geq 0.16$. This structure is more strongly distorted than that for $(P, T) = $ (\SI{125}{GPa}, \SI{300}{K}) in the classical MD simulation, where it has the $Fm\bar{3}m$ structure with symprec $=0.08$. 

\begin{table}[htbp]
  \caption{Results of the space group analysis for the structure obatined by MD and PIMD. Symprec is a threshold measuring the distortion of the crystal. The smaller symprec means that the space group of the crystal structure is analyzed more strictly. \label{pymatgen}}
  \centering
    \begin{tabular}{llccc}
    \hline
    \multirow{5}{*}{MD} & symprec & 0.08 & 0.12 & 0.16 \\
    \hline
    & 100 GPa 300 K& $P2_{1}/m$ & $R\bar{3}m$ &  $R\bar{3}m$ \\
    & 125 GPa 200 K& $C2/m$ & $R\bar{3}m$ & $R\bar{3}m$ \\
    & 125 GPa 300 K& $Fm\bar{3}m$ & $Fm\bar{3}m$ & $Fm\bar{3}m$ \\
    & 150 GPa 200 K& $R\bar{3}m$ & $R\bar{3}m$ & $R\bar{3}m$ \\
    & 150 GPa 300 K& $Fm\bar{3}m$ & $Fm\bar{3}m$ & $Fm\bar{3}m$ \\ \hline
    \hline
    \multirow{4}{*}{PIMD} & symprec & 0.08 & 0.12 & 0.16 \\
    \hline
    & 100 GPa 300 K& $P2_{1}$ & $P2_{1}$ & $Fm\bar{3}m$ \\
    & 125 GPa 200 K& $Fm\bar{3}m$ & $Fm\bar{3}m$ & $Fm\bar{3}m$ \\
    & 125 GPa 300 K& $Fm\bar{3}m$ & $Fm\bar{3}m$ & $Fm\bar{3}m$ \\
    & 150 GPa 200 K& $Fm\bar{3}m$ & $Fm\bar{3}m$ & $Fm\bar{3}m$ \\ \hline
    \end{tabular}
\end{table}

Comparing Figs.~\ref{md_xrd}, \ref{pimd_xrd} and Table~\ref{pymatgen}, we can clearly classify the calculated crystal structure. Let us first look into the results of the classical MD calculation. When the XRD pattern shows peak splitting at 2$\theta = 6.5^\circ$, the space group of the crystal structure is not $Fm\bar{3}m$, regardless of the value of symprec. For symprec larger than 0.12, the space group of the crystal structure is determined as $R\bar{3}m$. When smaller symprec is used, the crystal structures are identified as those having lower symmetries. For the space group of the crystal structures without peak splitting in the XRD pattern, it is always determined as $Fm\bar{3}m$ for all symprec $\geq 0.08$. Thus, we can safely conclude that the space group of the crystal structures at $(P, T) = $ (125 GPa, 300 K) and (150 GPa, 300 K) is $Fm\bar{3}m$ in the classical MD simulation.
On the other hand, we can see that the crystal structure at $(P, T) = $ (125 GPa, 200 K) and (150 GPa, 200 K) have a symmetry lower than $Fm\bar{3}m$. This result suggests that the temperature effect is critically important for the crystal symmetry at 125 GPa and 150 GPa. On the other hand, the space group of the crystal structure at $(P, T) = $ (100 GPa, 300 K) is not  $Fm\bar{3}m$, suggesting that the temperature effect of 300 K cannot stabilize the symmetric structure at 100 GPa.

Let us next move on to the space group analysis for the structures obtained by the PIMD simulation. The XRD patterns for the structure at $(P, T) = $ (\SI{100}{GPa}, \SI{300}{K}), (\SI{125}{GPa}, \SI{200}{K}), and  (\SI{150}{GPa}, \SI{200}{K})
show single peak at 2$\theta = 6.5^\circ$. The space group analysis in Table~\ref{pymatgen} also shows that all of these structures belong to the space group of $Fm\bar{3}m$ for symprec=0.16. However, the crystal structure at $(P, T) = $ (100 GPa, 300 K) is classified as $P2_1/m$ for smaller values of symprec, \hr{meaning that the stochastic error is sensitively detected.} This suggests that the point $(P, T) = $ (100 GPa, 300 K) resides close to the border of the $Fm\bar{3}m$ phase. It should be noted that there is recently an experimental report indicating that the necessary pressure to stabilize the $Fm\bar{3}m$ phase is lower than previously believed, which is consistent with the present calculation~\cite{sun_high-temperature_2021}.

\begin{figure}[htbp]
        \centering
        \includegraphics[keepaspectratio, scale=0.35]{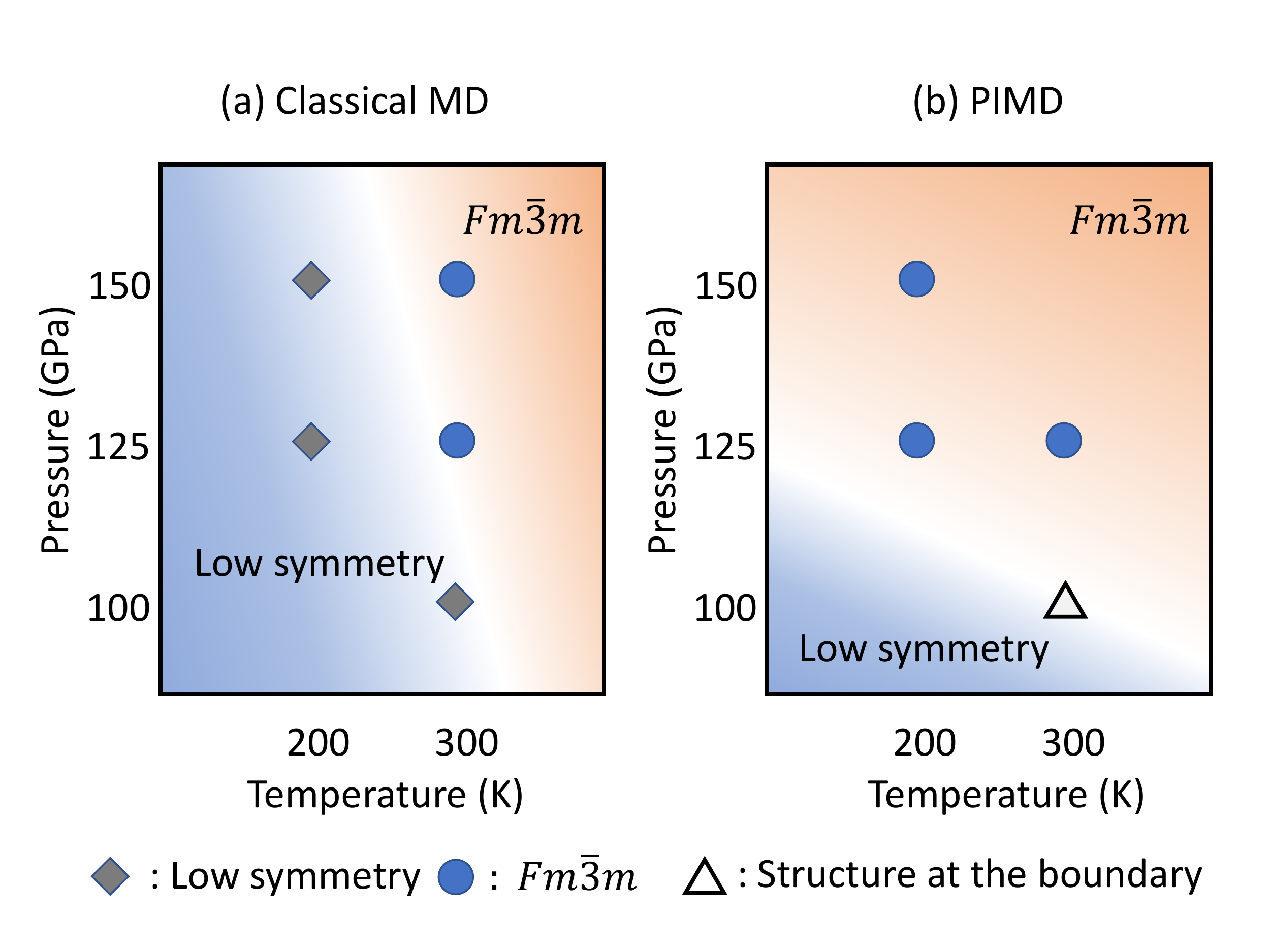}
        
        \caption{Phase diagram of the crystal structure of $\mathrm{LaH_{10}}$ crystal structure detemined by the classical MD (a) and PIMD (b). \label{phase}}
\end{figure}

Lastly, we compare the results of the MD and PIMD calculations. Given that the results for $(P, T) = $ (125 GPa, 200 K) and (150 GPa, 200 K) are $C2/m$ or $R\bar{3}m$ ($Fm\bar{3}m$) in the classical MD (PIMD) calculation, we can say that the $Fm\bar{3}m$ structure is stabilized by the quantum effect at temperature lower than 200 K when $P=125\sim 150$ GPa. This observation is consistent with the previous report based on the self-consistent harmonic approximation~\cite{errea_quantum_2020}. 
On the other hand, for $(P, T) = $ (125 GPa, 300 K) and (150 GPa, 300 K), both the MD and PIMD calculation give the $Fm\bar{3}m$ structure. This result suggests that the temperature effect can stabilize the highly symmetric $Fm\bar{3}m$ structure at a temperature higher than 300K. 
To visualize the present results, in Fig.~\ref{phase}, we show the phase diagram obtained by the classical MD and PIMD calculations. \hr{
A previous experiment shows that Tc abruptly drops when the crystal structure changes at lower pressures~\cite{sun_high-temperature_2021}. Given that the present symmetric structure is unstable for $P < 125$ GPa, this observation suggests that Tc is severely suppressed when the crystal structure is distorted. }

\section{conclusion}
We investigated the temperature and quantum effect on the crystal structure of $\mathrm{LaH_{10}}$. By classical MD and PIMD simulations, we determined the most stable crystal structures for 100 GPa $\leq P \leq$ 150 GPa and $T=$ 200 and 300 K. To analyze the symmetry of the calculated crystal structures, we calculated the XRD patterns and performed space group analysis. For 125 GPa and 150 GPa, we have shown that a temperature of 300 K has a crucial impact on stabilizing the symmetric $Fm\bar{3}m$ structure, which favors superconductivity.
While the $Fm\bar{3}m$ structure does not survive only by the temperature effect at 200 K, if we consider the quantum effect, the $Fm\bar{3}m$ structure becomes the most stable structure. We also found that the point of $(P, T)=$ (100 GPa, 300 K) locates close to the phase boundary between the $Fm\bar{3}m$ phase and those with lower symmetries.


\section*{Acknowledgements}
We would like to thank M. Shiga for the fruitful discussion. We acknowledge the Center for Computational Materials Science, Institute for Materials Research, Tohoku University for the use of MASAMUNE-IMR (MAterials science Supercomputing system for Advanced MUltiscale simulations towards NExt generation). (Project No. 202012-SCKXX-0009) This work was supported by a Grant-in-Aid for Scientific  Research (No. 19K14654, No. 19H05825, No. 20K21067, and No. 22H00110), “Program for Promoting Researches on the Supercomputer Fugaku” (Project ID: hp220166) from MEXT.

\bibliography{ref}

\begin{thebibliography}{64}%
\makeatletter
\providecommand \@ifxundefined [1]{%
 \@ifx{#1\undefined}
}%
\providecommand \@ifnum [1]{%
 \ifnum #1\expandafter \@firstoftwo
 \else \expandafter \@secondoftwo
 \fi
}%
\providecommand \@ifx [1]{%
 \ifx #1\expandafter \@firstoftwo
 \else \expandafter \@secondoftwo
 \fi
}%
\providecommand \natexlab [1]{#1}%
\providecommand \enquote  [1]{``#1''}%
\providecommand \bibnamefont  [1]{#1}%
\providecommand \bibfnamefont [1]{#1}%
\providecommand \citenamefont [1]{#1}%
\providecommand \href@noop [0]{\@secondoftwo}%
\providecommand \href [0]{\begingroup \@sanitize@url \@href}%
\providecommand \@href[1]{\@@startlink{#1}\@@href}%
\providecommand \@@href[1]{\endgroup#1\@@endlink}%
\providecommand \@sanitize@url [0]{\catcode `\\12\catcode `\$12\catcode
  `\&12\catcode `\#12\catcode `\^12\catcode `\_12\catcode `\%12\relax}%
\providecommand \@@startlink[1]{}%
\providecommand \@@endlink[0]{}%
\providecommand \url  [0]{\begingroup\@sanitize@url \@url }%
\providecommand \@url [1]{\endgroup\@href {#1}{\urlprefix }}%
\providecommand \urlprefix  [0]{URL }%
\providecommand \Eprint [0]{\href }%
\providecommand \doibase [0]{http://dx.doi.org/}%
\providecommand \selectlanguage [0]{\@gobble}%
\providecommand \bibinfo  [0]{\@secondoftwo}%
\providecommand \bibfield  [0]{\@secondoftwo}%
\providecommand \translation [1]{[#1]}%
\providecommand \BibitemOpen [0]{}%
\providecommand \bibitemStop [0]{}%
\providecommand \bibitemNoStop [0]{.\EOS\space}%
\providecommand \EOS [0]{\spacefactor3000\relax}%
\providecommand \BibitemShut  [1]{\csname bibitem#1\endcsname}%
\let\auto@bib@innerbib\@empty
\bibitem [{\citenamefont {Ashcroft}(1968)}]{ashcroft_metallic_1968.org}%
  \BibitemOpen
  \bibfield  {author} {\bibinfo {author} {\bibfnamefont {N.~W.}\ \bibnamefont
  {Ashcroft}},\ }\href {\doibase 10.1103/PhysRevLett.21.1748} {\bibfield
  {journal} {\bibinfo  {journal} {Physical Review Letters}\ }\textbf {\bibinfo
  {volume} {21}},\ \bibinfo {pages} {1748} (\bibinfo {year}
  {1968})}\BibitemShut {NoStop}%
\bibitem [{\citenamefont {Monacelli}\ \emph {et~al.}(2021)\citenamefont
  {Monacelli}, \citenamefont {Errea}, \citenamefont {Calandra},\ and\
  \citenamefont {Mauri}}]{monacelli_black_2021}%
  \BibitemOpen
  \bibfield  {author} {\bibinfo {author} {\bibfnamefont {L.}~\bibnamefont
  {Monacelli}}, \bibinfo {author} {\bibfnamefont {I.}~\bibnamefont {Errea}},
  \bibinfo {author} {\bibfnamefont {M.}~\bibnamefont {Calandra}}, \ and\
  \bibinfo {author} {\bibfnamefont {F.}~\bibnamefont {Mauri}},\ }\href
  {\doibase 10.1038/s41567-020-1009-3} {\bibfield  {journal} {\bibinfo
  {journal} {Nat. Phys.}\ }\textbf {\bibinfo {volume} {17}},\ \bibinfo {pages}
  {63} (\bibinfo {year} {2021})}\BibitemShut {NoStop}%
\bibitem [{\citenamefont {Eremets}\ and\ \citenamefont
  {Troyan}(2011)}]{eremets_conductive_2011}%
  \BibitemOpen
  \bibfield  {author} {\bibinfo {author} {\bibfnamefont {M.~I.}\ \bibnamefont
  {Eremets}}\ and\ \bibinfo {author} {\bibfnamefont {I.~A.}\ \bibnamefont
  {Troyan}},\ }\href {\doibase 10.1038/nmat3175} {\bibfield  {journal}
  {\bibinfo  {journal} {Nature Mater}\ }\textbf {\bibinfo {volume} {10}},\
  \bibinfo {pages} {927} (\bibinfo {year} {2011})}\BibitemShut {NoStop}%
\bibitem [{\citenamefont {Azadi}\ \emph {et~al.}(2014)\citenamefont {Azadi},
  \citenamefont {Monserrat}, \citenamefont {Foulkes},\ and\ \citenamefont
  {Needs}}]{azadi_dissociation_2014}%
  \BibitemOpen
  \bibfield  {author} {\bibinfo {author} {\bibfnamefont {S.}~\bibnamefont
  {Azadi}}, \bibinfo {author} {\bibfnamefont {B.}~\bibnamefont {Monserrat}},
  \bibinfo {author} {\bibfnamefont {W.}~\bibnamefont {Foulkes}}, \ and\
  \bibinfo {author} {\bibfnamefont {R.}~\bibnamefont {Needs}},\ }\href
  {\doibase 10.1103/PhysRevLett.112.165501} {\bibfield  {journal} {\bibinfo
  {journal} {Phys. Rev. Lett.}\ }\textbf {\bibinfo {volume} {112}},\ \bibinfo
  {pages} {165501} (\bibinfo {year} {2014})}\BibitemShut {NoStop}%
\bibitem [{\citenamefont {Cudazzo}\ \emph
  {et~al.}(2010{\natexlab{a}})\citenamefont {Cudazzo}, \citenamefont {Profeta},
  \citenamefont {Sanna}, \citenamefont {Floris}, \citenamefont {Continenza},
  \citenamefont {Massidda},\ and\ \citenamefont
  {Gross}}]{cudazzo_electron-phonon_2010}%
  \BibitemOpen
  \bibfield  {author} {\bibinfo {author} {\bibfnamefont {P.}~\bibnamefont
  {Cudazzo}}, \bibinfo {author} {\bibfnamefont {G.}~\bibnamefont {Profeta}},
  \bibinfo {author} {\bibfnamefont {A.}~\bibnamefont {Sanna}}, \bibinfo
  {author} {\bibfnamefont {A.}~\bibnamefont {Floris}}, \bibinfo {author}
  {\bibfnamefont {A.}~\bibnamefont {Continenza}}, \bibinfo {author}
  {\bibfnamefont {S.}~\bibnamefont {Massidda}}, \ and\ \bibinfo {author}
  {\bibfnamefont {E.~K.~U.}\ \bibnamefont {Gross}},\ }\href {\doibase
  10.1103/PhysRevB.81.134505} {\bibfield  {journal} {\bibinfo  {journal} {Phys.
  Rev. B}\ }\textbf {\bibinfo {volume} {81}},\ \bibinfo {pages} {134505}
  (\bibinfo {year} {2010}{\natexlab{a}})}\BibitemShut {NoStop}%
\bibitem [{\citenamefont {Cudazzo}\ \emph
  {et~al.}(2010{\natexlab{b}})\citenamefont {Cudazzo}, \citenamefont {Profeta},
  \citenamefont {Sanna}, \citenamefont {Floris}, \citenamefont {Continenza},
  \citenamefont {Massidda},\ and\ \citenamefont
  {Gross}}]{cudazzo_electron-phonon_2010-1}%
  \BibitemOpen
  \bibfield  {author} {\bibinfo {author} {\bibfnamefont {P.}~\bibnamefont
  {Cudazzo}}, \bibinfo {author} {\bibfnamefont {G.}~\bibnamefont {Profeta}},
  \bibinfo {author} {\bibfnamefont {A.}~\bibnamefont {Sanna}}, \bibinfo
  {author} {\bibfnamefont {A.}~\bibnamefont {Floris}}, \bibinfo {author}
  {\bibfnamefont {A.}~\bibnamefont {Continenza}}, \bibinfo {author}
  {\bibfnamefont {S.}~\bibnamefont {Massidda}}, \ and\ \bibinfo {author}
  {\bibfnamefont {E.~K.~U.}\ \bibnamefont {Gross}},\ }\href {\doibase
  10.1103/PhysRevB.81.134506} {\bibfield  {journal} {\bibinfo  {journal} {Phys.
  Rev. B}\ }\textbf {\bibinfo {volume} {81}},\ \bibinfo {pages} {134506}
  (\bibinfo {year} {2010}{\natexlab{b}})}\BibitemShut {NoStop}%
\bibitem [{\citenamefont {Dalladay-Simpson}\ \emph {et~al.}(2016)\citenamefont
  {Dalladay-Simpson}, \citenamefont {Howie},\ and\ \citenamefont
  {Gregoryanz}}]{dalladay-simpson_evidence_2016}%
  \BibitemOpen
  \bibfield  {author} {\bibinfo {author} {\bibfnamefont {P.}~\bibnamefont
  {Dalladay-Simpson}}, \bibinfo {author} {\bibfnamefont {R.~T.}\ \bibnamefont
  {Howie}}, \ and\ \bibinfo {author} {\bibfnamefont {E.}~\bibnamefont
  {Gregoryanz}},\ }\href {\doibase 10.1038/nature16164} {\bibfield  {journal}
  {\bibinfo  {journal} {Nature}\ }\textbf {\bibinfo {volume} {529}},\ \bibinfo
  {pages} {63} (\bibinfo {year} {2016})}\BibitemShut {NoStop}%
\bibitem [{\citenamefont {McMinis}\ \emph {et~al.}(2015)\citenamefont
  {McMinis}, \citenamefont {Clay}, \citenamefont {Lee},\ and\ \citenamefont
  {Morales}}]{mcminis_molecular_2015}%
  \BibitemOpen
  \bibfield  {author} {\bibinfo {author} {\bibfnamefont {J.}~\bibnamefont
  {McMinis}}, \bibinfo {author} {\bibfnamefont {R.~C.}\ \bibnamefont {Clay}},
  \bibinfo {author} {\bibfnamefont {D.}~\bibnamefont {Lee}}, \ and\ \bibinfo
  {author} {\bibfnamefont {M.~A.}\ \bibnamefont {Morales}},\ }\href {\doibase
  10.1103/PhysRevLett.114.105305} {\bibfield  {journal} {\bibinfo  {journal}
  {Phys. Rev. Lett.}\ }\textbf {\bibinfo {volume} {114}},\ \bibinfo {pages}
  {105305} (\bibinfo {year} {2015})}\BibitemShut {NoStop}%
\bibitem [{\citenamefont {Dias}\ and\ \citenamefont
  {Silvera}(2017)}]{dias_observation_2017}%
  \BibitemOpen
  \bibfield  {author} {\bibinfo {author} {\bibfnamefont {R.~P.}\ \bibnamefont
  {Dias}}\ and\ \bibinfo {author} {\bibfnamefont {I.~F.}\ \bibnamefont
  {Silvera}},\ }\href {\doibase 10.1126/science.aal1579} {\bibfield  {journal}
  {\bibinfo  {journal} {Science}\ }\textbf {\bibinfo {volume} {355}},\ \bibinfo
  {pages} {715} (\bibinfo {year} {2017})}\BibitemShut {NoStop}%
\bibitem [{\citenamefont {Loubeyre}\ \emph {et~al.}(2020)\citenamefont
  {Loubeyre}, \citenamefont {Occelli},\ and\ \citenamefont
  {Dumas}}]{Loubeyre_2020}%
  \BibitemOpen
  \bibfield  {author} {\bibinfo {author} {\bibfnamefont {P.}~\bibnamefont
  {Loubeyre}}, \bibinfo {author} {\bibfnamefont {F.}~\bibnamefont {Occelli}}, \
  and\ \bibinfo {author} {\bibfnamefont {P.}~\bibnamefont {Dumas}},\
  }\href@noop {} {\bibfield  {journal} {\bibinfo  {journal} {Nature}\ }\textbf
  {\bibinfo {volume} {577}},\ \bibinfo {pages} {631} (\bibinfo {year}
  {2020})}\BibitemShut {NoStop}%
\bibitem [{\citenamefont {McMahon}\ \emph {et~al.}(2012)\citenamefont
  {McMahon}, \citenamefont {Morales}, \citenamefont {Pierleoni},\ and\
  \citenamefont {Ceperley}}]{mcmahon_properties_2012}%
  \BibitemOpen
  \bibfield  {author} {\bibinfo {author} {\bibfnamefont {J.~M.}\ \bibnamefont
  {McMahon}}, \bibinfo {author} {\bibfnamefont {M.~A.}\ \bibnamefont
  {Morales}}, \bibinfo {author} {\bibfnamefont {C.}~\bibnamefont {Pierleoni}},
  \ and\ \bibinfo {author} {\bibfnamefont {D.~M.}\ \bibnamefont {Ceperley}},\
  }\href {\doibase 10.1103/RevModPhys.84.1607} {\bibfield  {journal} {\bibinfo
  {journal} {Rev. Mod. Phys.}\ }\textbf {\bibinfo {volume} {84}},\ \bibinfo
  {pages} {1607} (\bibinfo {year} {2012})}\BibitemShut {NoStop}%
\bibitem [{\citenamefont {Gilman}(1971)}]{gilman_lithium_1971}%
  \BibitemOpen
  \bibfield  {author} {\bibinfo {author} {\bibfnamefont {J.~J.}\ \bibnamefont
  {Gilman}},\ }\href {\doibase 10.1103/PhysRevLett.26.546} {\bibfield
  {journal} {\bibinfo  {journal} {Phys. Rev. Lett.}\ }\textbf {\bibinfo
  {volume} {26}},\ \bibinfo {pages} {546} (\bibinfo {year} {1971})}\BibitemShut
  {NoStop}%
\bibitem [{\citenamefont {Ashcroft}(2004)}]{ashcroft_hydrogen_2004}%
  \BibitemOpen
  \bibfield  {author} {\bibinfo {author} {\bibfnamefont {N.~W.}\ \bibnamefont
  {Ashcroft}},\ }\href {\doibase 10.1103/PhysRevLett.92.187002} {\bibfield
  {journal} {\bibinfo  {journal} {Phys. Rev. Lett.}\ }\textbf {\bibinfo
  {volume} {92}},\ \bibinfo {pages} {187002} (\bibinfo {year}
  {2004})}\BibitemShut {NoStop}%
\bibitem [{\citenamefont {Flores-Livas}\ \emph {et~al.}(2020)\citenamefont
  {Flores-Livas}, \citenamefont {Boeri}, \citenamefont {Sanna}, \citenamefont
  {Profeta}, \citenamefont {Arita},\ and\ \citenamefont
  {Eremets}}]{flores-livas_perspective_2020}%
  \BibitemOpen
  \bibfield  {author} {\bibinfo {author} {\bibfnamefont {J.~A.}\ \bibnamefont
  {Flores-Livas}}, \bibinfo {author} {\bibfnamefont {L.}~\bibnamefont {Boeri}},
  \bibinfo {author} {\bibfnamefont {A.}~\bibnamefont {Sanna}}, \bibinfo
  {author} {\bibfnamefont {G.}~\bibnamefont {Profeta}}, \bibinfo {author}
  {\bibfnamefont {R.}~\bibnamefont {Arita}}, \ and\ \bibinfo {author}
  {\bibfnamefont {M.}~\bibnamefont {Eremets}},\ }\href@noop {} {\bibfield
  {journal} {\bibinfo  {journal} {Physics Reports}\ ,\ \bibinfo {pages} {78}}
  (\bibinfo {year} {2020})}\BibitemShut {NoStop}%
\bibitem [{\citenamefont {Quan}\ \emph {et~al.}(2019)\citenamefont {Quan},
  \citenamefont {Ghosh},\ and\ \citenamefont {Pickett}}]{quan_compressed_2019}%
  \BibitemOpen
  \bibfield  {author} {\bibinfo {author} {\bibfnamefont {Y.}~\bibnamefont
  {Quan}}, \bibinfo {author} {\bibfnamefont {S.~S.}\ \bibnamefont {Ghosh}}, \
  and\ \bibinfo {author} {\bibfnamefont {W.~E.}\ \bibnamefont {Pickett}},\
  }\href {\doibase 10.1103/PhysRevB.100.184505} {\bibfield  {journal} {\bibinfo
   {journal} {Phys. Rev. B}\ }\textbf {\bibinfo {volume} {100}},\ \bibinfo
  {pages} {184505} (\bibinfo {year} {2019})}\BibitemShut {NoStop}%
\bibitem [{\citenamefont {Feng}\ \emph {et~al.}(2015)\citenamefont {Feng},
  \citenamefont {Zhang}, \citenamefont {Gao}, \citenamefont {Liu},\ and\
  \citenamefont {Wang}}]{feng_compressed_2015}%
  \BibitemOpen
  \bibfield  {author} {\bibinfo {author} {\bibfnamefont {X.}~\bibnamefont
  {Feng}}, \bibinfo {author} {\bibfnamefont {J.}~\bibnamefont {Zhang}},
  \bibinfo {author} {\bibfnamefont {G.}~\bibnamefont {Gao}}, \bibinfo {author}
  {\bibfnamefont {H.}~\bibnamefont {Liu}}, \ and\ \bibinfo {author}
  {\bibfnamefont {H.}~\bibnamefont {Wang}},\ }\href {\doibase
  10.1039/C5RA11459D} {\bibfield  {journal} {\bibinfo  {journal} {RSC Adv.}\
  }\textbf {\bibinfo {volume} {5}},\ \bibinfo {pages} {59292} (\bibinfo {year}
  {2015})}\BibitemShut {NoStop}%
\bibitem [{\citenamefont {Ge}\ \emph {et~al.}(2016)\citenamefont {Ge},
  \citenamefont {Zhang},\ and\ \citenamefont {Yao}}]{ge_first-principles_2016}%
  \BibitemOpen
  \bibfield  {author} {\bibinfo {author} {\bibfnamefont {Y.}~\bibnamefont
  {Ge}}, \bibinfo {author} {\bibfnamefont {F.}~\bibnamefont {Zhang}}, \ and\
  \bibinfo {author} {\bibfnamefont {Y.}~\bibnamefont {Yao}},\ }\href {\doibase
  10.1103/PhysRevB.93.224513} {\bibfield  {journal} {\bibinfo  {journal} {Phys.
  Rev. B}\ }\textbf {\bibinfo {volume} {93}},\ \bibinfo {pages} {224513}
  (\bibinfo {year} {2016})}\BibitemShut {NoStop}%
\bibitem [{\citenamefont {Errea}\ \emph {et~al.}(2015)\citenamefont {Errea},
  \citenamefont {Calandra}, \citenamefont {Pickard}, \citenamefont {Nelson},
  \citenamefont {Needs}, \citenamefont {Li}, \citenamefont {Liu}, \citenamefont
  {Zhang}, \citenamefont {Ma},\ and\ \citenamefont
  {Mauri}}]{errea_high-pressure_2015}%
  \BibitemOpen
  \bibfield  {author} {\bibinfo {author} {\bibfnamefont {I.}~\bibnamefont
  {Errea}}, \bibinfo {author} {\bibfnamefont {M.}~\bibnamefont {Calandra}},
  \bibinfo {author} {\bibfnamefont {C.~J.}\ \bibnamefont {Pickard}}, \bibinfo
  {author} {\bibfnamefont {J.}~\bibnamefont {Nelson}}, \bibinfo {author}
  {\bibfnamefont {R.~J.}\ \bibnamefont {Needs}}, \bibinfo {author}
  {\bibfnamefont {Y.}~\bibnamefont {Li}}, \bibinfo {author} {\bibfnamefont
  {H.}~\bibnamefont {Liu}}, \bibinfo {author} {\bibfnamefont {Y.}~\bibnamefont
  {Zhang}}, \bibinfo {author} {\bibfnamefont {Y.}~\bibnamefont {Ma}}, \ and\
  \bibinfo {author} {\bibfnamefont {F.}~\bibnamefont {Mauri}},\ }\href
  {\doibase 10.1103/PhysRevLett.114.157004} {\bibfield  {journal} {\bibinfo
  {journal} {Phys. Rev. Lett.}\ }\textbf {\bibinfo {volume} {114}},\ \bibinfo
  {pages} {157004} (\bibinfo {year} {2015})}\BibitemShut {NoStop}%
\bibitem [{\citenamefont {Zurek}\ and\ \citenamefont
  {Bi}(2019)}]{zurek_high-temperature_2019}%
  \BibitemOpen
  \bibfield  {author} {\bibinfo {author} {\bibfnamefont {E.}~\bibnamefont
  {Zurek}}\ and\ \bibinfo {author} {\bibfnamefont {T.}~\bibnamefont {Bi}},\
  }\href {\doibase 10.1063/1.5079225} {\bibfield  {journal} {\bibinfo
  {journal} {J. Chem. Phys.}\ }\textbf {\bibinfo {volume} {150}},\ \bibinfo
  {pages} {050901} (\bibinfo {year} {2019})}\BibitemShut {NoStop}%
\bibitem [{\citenamefont {Papaconstantopoulos}\ \emph
  {et~al.}(2020)\citenamefont {Papaconstantopoulos}, \citenamefont {Mehl},\
  and\ \citenamefont {Chang}}]{papaconstantopoulos_high-temperature_2020}%
  \BibitemOpen
  \bibfield  {author} {\bibinfo {author} {\bibfnamefont {D.~A.}\ \bibnamefont
  {Papaconstantopoulos}}, \bibinfo {author} {\bibfnamefont {M.~J.}\
  \bibnamefont {Mehl}}, \ and\ \bibinfo {author} {\bibfnamefont {P.-H.}\
  \bibnamefont {Chang}},\ }\href {\doibase 10.1103/PhysRevB.101.060506}
  {\bibfield  {journal} {\bibinfo  {journal} {Phys. Rev. B}\ }\textbf {\bibinfo
  {volume} {101}},\ \bibinfo {pages} {060506} (\bibinfo {year}
  {2020})}\BibitemShut {NoStop}%
\bibitem [{\citenamefont {Peng}\ \emph {et~al.}(2017)\citenamefont {Peng},
  \citenamefont {Sun}, \citenamefont {Pickard}, \citenamefont {Needs},
  \citenamefont {Wu},\ and\ \citenamefont {Ma}}]{peng_hydrogen_2017}%
  \BibitemOpen
  \bibfield  {author} {\bibinfo {author} {\bibfnamefont {F.}~\bibnamefont
  {Peng}}, \bibinfo {author} {\bibfnamefont {Y.}~\bibnamefont {Sun}}, \bibinfo
  {author} {\bibfnamefont {C.~J.}\ \bibnamefont {Pickard}}, \bibinfo {author}
  {\bibfnamefont {R.~J.}\ \bibnamefont {Needs}}, \bibinfo {author}
  {\bibfnamefont {Q.}~\bibnamefont {Wu}}, \ and\ \bibinfo {author}
  {\bibfnamefont {Y.}~\bibnamefont {Ma}},\ }\href {\doibase
  10.1103/PhysRevLett.119.107001} {\bibfield  {journal} {\bibinfo  {journal}
  {Phys. Rev. Lett.}\ }\textbf {\bibinfo {volume} {119}},\ \bibinfo {pages}
  {107001} (\bibinfo {year} {2017})}\BibitemShut {NoStop}%
\bibitem [{\citenamefont {Di~Cataldo}\ \emph {et~al.}(2021)\citenamefont
  {Di~Cataldo}, \citenamefont {Heil}, \citenamefont {von~der Linden},\ and\
  \citenamefont {Boeri}}]{di_cataldo_bh_2021}%
  \BibitemOpen
  \bibfield  {author} {\bibinfo {author} {\bibfnamefont {S.}~\bibnamefont
  {Di~Cataldo}}, \bibinfo {author} {\bibfnamefont {C.}~\bibnamefont {Heil}},
  \bibinfo {author} {\bibfnamefont {W.}~\bibnamefont {von~der Linden}}, \ and\
  \bibinfo {author} {\bibfnamefont {L.}~\bibnamefont {Boeri}},\ }\href
  {\doibase 10.1103/PhysRevB.104.L020511} {\bibfield  {journal} {\bibinfo
  {journal} {Phys. Rev. B}\ }\textbf {\bibinfo {volume} {104}},\ \bibinfo
  {pages} {L020511} (\bibinfo {year} {2021})}\BibitemShut {NoStop}%
\bibitem [{\citenamefont {Liu}\ \emph {et~al.}(2019)\citenamefont {Liu},
  \citenamefont {Wang}, \citenamefont {Yi}, \citenamefont {Kim}, \citenamefont
  {Kim},\ and\ \citenamefont {Cho}}]{liu_microscopic_2019}%
  \BibitemOpen
  \bibfield  {author} {\bibinfo {author} {\bibfnamefont {L.}~\bibnamefont
  {Liu}}, \bibinfo {author} {\bibfnamefont {C.}~\bibnamefont {Wang}}, \bibinfo
  {author} {\bibfnamefont {S.}~\bibnamefont {Yi}}, \bibinfo {author}
  {\bibfnamefont {K.~W.}\ \bibnamefont {Kim}}, \bibinfo {author} {\bibfnamefont
  {J.}~\bibnamefont {Kim}}, \ and\ \bibinfo {author} {\bibfnamefont {J.-H.}\
  \bibnamefont {Cho}},\ }\href {\doibase 10.1103/PhysRevB.99.140501} {\bibfield
   {journal} {\bibinfo  {journal} {Phys. Rev. B}\ }\textbf {\bibinfo {volume}
  {99}},\ \bibinfo {pages} {140501} (\bibinfo {year} {2019})}\BibitemShut
  {NoStop}%
\bibitem [{\citenamefont {Gao}\ \emph {et~al.}(2021)\citenamefont {Gao},
  \citenamefont {Yan}, \citenamefont {Lu},\ and\ \citenamefont
  {Xiang}}]{gao_phonon-mediated_2021}%
  \BibitemOpen
  \bibfield  {author} {\bibinfo {author} {\bibfnamefont {M.}~\bibnamefont
  {Gao}}, \bibinfo {author} {\bibfnamefont {X.-W.}\ \bibnamefont {Yan}},
  \bibinfo {author} {\bibfnamefont {Z.-Y.}\ \bibnamefont {Lu}}, \ and\ \bibinfo
  {author} {\bibfnamefont {T.}~\bibnamefont {Xiang}},\ }\href {\doibase
  10.1103/PhysRevB.104.L100504} {\bibfield  {journal} {\bibinfo  {journal}
  {Phys. Rev. B}\ }\textbf {\bibinfo {volume} {104}},\ \bibinfo {pages}
  {L100504} (\bibinfo {year} {2021})}\BibitemShut {NoStop}%
\bibitem [{\citenamefont {Liu}\ \emph {et~al.}(2017)\citenamefont {Liu},
  \citenamefont {Naumov}, \citenamefont {Hoffmann}, \citenamefont {Ashcroft},\
  and\ \citenamefont {Hemley}}]{liu_potential_2017}%
  \BibitemOpen
  \bibfield  {author} {\bibinfo {author} {\bibfnamefont {H.}~\bibnamefont
  {Liu}}, \bibinfo {author} {\bibfnamefont {I.~I.}\ \bibnamefont {Naumov}},
  \bibinfo {author} {\bibfnamefont {R.}~\bibnamefont {Hoffmann}}, \bibinfo
  {author} {\bibfnamefont {N.~W.}\ \bibnamefont {Ashcroft}}, \ and\ \bibinfo
  {author} {\bibfnamefont {R.~J.}\ \bibnamefont {Hemley}},\ }\href {\doibase
  10.1073/pnas.1704505114} {\bibfield  {journal} {\bibinfo  {journal} {Proc
  Natl Acad Sci USA}\ }\textbf {\bibinfo {volume} {114}},\ \bibinfo {pages}
  {6990} (\bibinfo {year} {2017})}\BibitemShut {NoStop}%
\bibitem [{\citenamefont {Shipley}\ \emph {et~al.}(2020)\citenamefont
  {Shipley}, \citenamefont {Hutcheon}, \citenamefont {Johnson}, \citenamefont
  {Needs},\ and\ \citenamefont {Pickard}}]{shipley_stability_2020}%
  \BibitemOpen
  \bibfield  {author} {\bibinfo {author} {\bibfnamefont {A.~M.}\ \bibnamefont
  {Shipley}}, \bibinfo {author} {\bibfnamefont {M.~J.}\ \bibnamefont
  {Hutcheon}}, \bibinfo {author} {\bibfnamefont {M.~S.}\ \bibnamefont
  {Johnson}}, \bibinfo {author} {\bibfnamefont {R.~J.}\ \bibnamefont {Needs}},
  \ and\ \bibinfo {author} {\bibfnamefont {C.~J.}\ \bibnamefont {Pickard}},\
  }\href {\doibase 10.1103/PhysRevB.101.224511} {\bibfield  {journal} {\bibinfo
   {journal} {Phys. Rev. B}\ }\textbf {\bibinfo {volume} {101}},\ \bibinfo
  {pages} {224511} (\bibinfo {year} {2020})}\BibitemShut {NoStop}%
\bibitem [{\citenamefont {Kim}\ \emph {et~al.}(2009)\citenamefont {Kim},
  \citenamefont {Scheicher},\ and\ \citenamefont {Ahuja}}]{kim_predicted_2009}%
  \BibitemOpen
  \bibfield  {author} {\bibinfo {author} {\bibfnamefont {D.~Y.}\ \bibnamefont
  {Kim}}, \bibinfo {author} {\bibfnamefont {R.~H.}\ \bibnamefont {Scheicher}},
  \ and\ \bibinfo {author} {\bibfnamefont {R.}~\bibnamefont {Ahuja}},\ }\href
  {\doibase 10.1103/PhysRevLett.103.077002} {\bibfield  {journal} {\bibinfo
  {journal} {Phys. Rev. Lett.}\ }\textbf {\bibinfo {volume} {103}},\ \bibinfo
  {pages} {077002} (\bibinfo {year} {2009})}\BibitemShut {NoStop}%
\bibitem [{\citenamefont {Verma}\ \emph {et~al.}(2021)\citenamefont {Verma},
  \citenamefont {Modak}, \citenamefont {Schrodi}, \citenamefont {Aperis},\ and\
  \citenamefont {Oppeneer}}]{verma_prediction_2021}%
  \BibitemOpen
  \bibfield  {author} {\bibinfo {author} {\bibfnamefont {A.~K.}\ \bibnamefont
  {Verma}}, \bibinfo {author} {\bibfnamefont {P.}~\bibnamefont {Modak}},
  \bibinfo {author} {\bibfnamefont {F.}~\bibnamefont {Schrodi}}, \bibinfo
  {author} {\bibfnamefont {A.}~\bibnamefont {Aperis}}, \ and\ \bibinfo {author}
  {\bibfnamefont {P.~M.}\ \bibnamefont {Oppeneer}},\ }\href {\doibase
  10.1103/PhysRevB.104.174506} {\bibfield  {journal} {\bibinfo  {journal}
  {Phys. Rev. B}\ }\textbf {\bibinfo {volume} {104}},\ \bibinfo {pages}
  {174506} (\bibinfo {year} {2021})}\BibitemShut {NoStop}%
\bibitem [{\citenamefont {Li}\ \emph {et~al.}(2015)\citenamefont {Li},
  \citenamefont {Hao}, \citenamefont {Liu}, \citenamefont {Tse}, \citenamefont
  {Wang},\ and\ \citenamefont {Ma}}]{li_pressure-stabilized_2015}%
  \BibitemOpen
  \bibfield  {author} {\bibinfo {author} {\bibfnamefont {Y.}~\bibnamefont
  {Li}}, \bibinfo {author} {\bibfnamefont {J.}~\bibnamefont {Hao}}, \bibinfo
  {author} {\bibfnamefont {H.}~\bibnamefont {Liu}}, \bibinfo {author}
  {\bibfnamefont {J.~S.}\ \bibnamefont {Tse}}, \bibinfo {author} {\bibfnamefont
  {Y.}~\bibnamefont {Wang}}, \ and\ \bibinfo {author} {\bibfnamefont
  {Y.}~\bibnamefont {Ma}},\ }\href {\doibase 10.1038/srep09948} {\bibfield
  {journal} {\bibinfo  {journal} {Sci Rep}\ }\textbf {\bibinfo {volume} {5}},\
  \bibinfo {pages} {9948} (\bibinfo {year} {2015})}\BibitemShut {NoStop}%
\bibitem [{\citenamefont {Sun}\ \emph {et~al.}(2019)\citenamefont {Sun},
  \citenamefont {Lv}, \citenamefont {Xie}, \citenamefont {Liu},\ and\
  \citenamefont {Ma}}]{sun_route_2019}%
  \BibitemOpen
  \bibfield  {author} {\bibinfo {author} {\bibfnamefont {Y.}~\bibnamefont
  {Sun}}, \bibinfo {author} {\bibfnamefont {J.}~\bibnamefont {Lv}}, \bibinfo
  {author} {\bibfnamefont {Y.}~\bibnamefont {Xie}}, \bibinfo {author}
  {\bibfnamefont {H.}~\bibnamefont {Liu}}, \ and\ \bibinfo {author}
  {\bibfnamefont {Y.}~\bibnamefont {Ma}},\ }\href {\doibase
  10.1103/PhysRevLett.123.097001} {\bibfield  {journal} {\bibinfo  {journal}
  {Phys. Rev. Lett.}\ }\textbf {\bibinfo {volume} {123}},\ \bibinfo {pages}
  {097001} (\bibinfo {year} {2019})}\BibitemShut {NoStop}%
\bibitem [{\citenamefont {Yi}\ \emph {et~al.}(2021)\citenamefont {Yi},
  \citenamefont {Wang}, \citenamefont {Jeon},\ and\ \citenamefont
  {Cho}}]{yi_stability_2021}%
  \BibitemOpen
  \bibfield  {author} {\bibinfo {author} {\bibfnamefont {S.}~\bibnamefont
  {Yi}}, \bibinfo {author} {\bibfnamefont {C.}~\bibnamefont {Wang}}, \bibinfo
  {author} {\bibfnamefont {H.}~\bibnamefont {Jeon}}, \ and\ \bibinfo {author}
  {\bibfnamefont {J.-H.}\ \bibnamefont {Cho}},\ }\href {\doibase
  10.1103/PhysRevMaterials.5.024801} {\bibfield  {journal} {\bibinfo  {journal}
  {Phys. Rev. Materials}\ }\textbf {\bibinfo {volume} {5}},\ \bibinfo {pages}
  {024801} (\bibinfo {year} {2021})}\BibitemShut {NoStop}%
\bibitem [{\citenamefont {Kruglov}\ \emph {et~al.}(2020)\citenamefont
  {Kruglov}, \citenamefont {Semenok}, \citenamefont {Song}, \citenamefont
  {Szcze\'{s}niak}, \citenamefont {Wrona}, \citenamefont {Akashi},
  \citenamefont {Davari~Esfahani}, \citenamefont {Duan}, \citenamefont {Cui},
  \citenamefont {Kvashnin},\ and\ \citenamefont
  {Oganov}}]{kruglov_superconductivity_2020}%
  \BibitemOpen
  \bibfield  {author} {\bibinfo {author} {\bibfnamefont {I.~A.}\ \bibnamefont
  {Kruglov}}, \bibinfo {author} {\bibfnamefont {D.~V.}\ \bibnamefont
  {Semenok}}, \bibinfo {author} {\bibfnamefont {H.}~\bibnamefont {Song}},
  \bibinfo {author} {\bibfnamefont {R.}~\bibnamefont {Szcze\'{s}niak}},
  \bibinfo {author} {\bibfnamefont {I.~A.}\ \bibnamefont {Wrona}}, \bibinfo
  {author} {\bibfnamefont {R.}~\bibnamefont {Akashi}}, \bibinfo {author}
  {\bibfnamefont {M.~M.}\ \bibnamefont {Davari~Esfahani}}, \bibinfo {author}
  {\bibfnamefont {D.}~\bibnamefont {Duan}}, \bibinfo {author} {\bibfnamefont
  {T.}~\bibnamefont {Cui}}, \bibinfo {author} {\bibfnamefont {A.~G.}\
  \bibnamefont {Kvashnin}}, \ and\ \bibinfo {author} {\bibfnamefont {A.~R.}\
  \bibnamefont {Oganov}},\ }\href {\doibase 10.1103/PhysRevB.101.024508}
  {\bibfield  {journal} {\bibinfo  {journal} {Phys. Rev. B}\ }\textbf {\bibinfo
  {volume} {101}},\ \bibinfo {pages} {024508} (\bibinfo {year}
  {2020})}\BibitemShut {NoStop}%
\bibitem [{\citenamefont {Duan}\ \emph {et~al.}(2015)\citenamefont {Duan},
  \citenamefont {Liu}, \citenamefont {Tian}, \citenamefont {Li}, \citenamefont
  {Huang}, \citenamefont {Zhao}, \citenamefont {Yu}, \citenamefont {Liu},
  \citenamefont {Tian},\ and\ \citenamefont
  {Cui}}]{duan_pressure-induced_2015}%
  \BibitemOpen
  \bibfield  {author} {\bibinfo {author} {\bibfnamefont {D.}~\bibnamefont
  {Duan}}, \bibinfo {author} {\bibfnamefont {Y.}~\bibnamefont {Liu}}, \bibinfo
  {author} {\bibfnamefont {F.}~\bibnamefont {Tian}}, \bibinfo {author}
  {\bibfnamefont {D.}~\bibnamefont {Li}}, \bibinfo {author} {\bibfnamefont
  {X.}~\bibnamefont {Huang}}, \bibinfo {author} {\bibfnamefont
  {Z.}~\bibnamefont {Zhao}}, \bibinfo {author} {\bibfnamefont {H.}~\bibnamefont
  {Yu}}, \bibinfo {author} {\bibfnamefont {B.}~\bibnamefont {Liu}}, \bibinfo
  {author} {\bibfnamefont {W.}~\bibnamefont {Tian}}, \ and\ \bibinfo {author}
  {\bibfnamefont {T.}~\bibnamefont {Cui}},\ }\href {\doibase 10.1038/srep06968}
  {\bibfield  {journal} {\bibinfo  {journal} {Sci Rep}\ }\textbf {\bibinfo
  {volume} {4}},\ \bibinfo {pages} {6968} (\bibinfo {year} {2015})}\BibitemShut
  {NoStop}%
\bibitem [{\citenamefont {Troyan}\ \emph {et~al.}(2021)\citenamefont {Troyan},
  \citenamefont {Semenok}, \citenamefont {Kvashnin}, \citenamefont {Sadakov},
  \citenamefont {Sobolevskiy}, \citenamefont {Pudalov}, \citenamefont
  {Ivanova}, \citenamefont {Prakapenka}, \citenamefont {Greenberg},
  \citenamefont {Gavriliuk}, \citenamefont {Lyubutin}, \citenamefont
  {Struzhkin}, \citenamefont {Bergara}, \citenamefont {Errea}, \citenamefont
  {Bianco}, \citenamefont {Calandra}, \citenamefont {Mauri}, \citenamefont
  {Monacelli}, \citenamefont {Akashi},\ and\ \citenamefont
  {Oganov}}]{troyan_anomalous_2021}%
  \BibitemOpen
  \bibfield  {author} {\bibinfo {author} {\bibfnamefont {I.~A.}\ \bibnamefont
  {Troyan}}, \bibinfo {author} {\bibfnamefont {D.~V.}\ \bibnamefont {Semenok}},
  \bibinfo {author} {\bibfnamefont {A.~G.}\ \bibnamefont {Kvashnin}}, \bibinfo
  {author} {\bibfnamefont {A.~V.}\ \bibnamefont {Sadakov}}, \bibinfo {author}
  {\bibfnamefont {O.~A.}\ \bibnamefont {Sobolevskiy}}, \bibinfo {author}
  {\bibfnamefont {V.~M.}\ \bibnamefont {Pudalov}}, \bibinfo {author}
  {\bibfnamefont {A.~G.}\ \bibnamefont {Ivanova}}, \bibinfo {author}
  {\bibfnamefont {V.~B.}\ \bibnamefont {Prakapenka}}, \bibinfo {author}
  {\bibfnamefont {E.}~\bibnamefont {Greenberg}}, \bibinfo {author}
  {\bibfnamefont {A.~G.}\ \bibnamefont {Gavriliuk}}, \bibinfo {author}
  {\bibfnamefont {I.~S.}\ \bibnamefont {Lyubutin}}, \bibinfo {author}
  {\bibfnamefont {V.~V.}\ \bibnamefont {Struzhkin}}, \bibinfo {author}
  {\bibfnamefont {A.}~\bibnamefont {Bergara}}, \bibinfo {author} {\bibfnamefont
  {I.}~\bibnamefont {Errea}}, \bibinfo {author} {\bibfnamefont
  {R.}~\bibnamefont {Bianco}}, \bibinfo {author} {\bibfnamefont
  {M.}~\bibnamefont {Calandra}}, \bibinfo {author} {\bibfnamefont
  {F.}~\bibnamefont {Mauri}}, \bibinfo {author} {\bibfnamefont
  {L.}~\bibnamefont {Monacelli}}, \bibinfo {author} {\bibfnamefont
  {R.}~\bibnamefont {Akashi}}, \ and\ \bibinfo {author} {\bibfnamefont {A.~R.}\
  \bibnamefont {Oganov}},\ }\href {\doibase 10.1002/adma.202006832} {\bibfield
  {journal} {\bibinfo  {journal} {Adv. Mater.}\ }\textbf {\bibinfo {volume}
  {33}},\ \bibinfo {pages} {2006832} (\bibinfo {year} {2021})}\BibitemShut
  {NoStop}%
\bibitem [{\citenamefont {Drozdov}\ \emph {et~al.}(2015)\citenamefont
  {Drozdov}, \citenamefont {Eremets}, \citenamefont {Troyan}, \citenamefont
  {Ksenofontov},\ and\ \citenamefont {Shylin}}]{drozdov_conventional_2015}%
  \BibitemOpen
  \bibfield  {author} {\bibinfo {author} {\bibfnamefont {A.~P.}\ \bibnamefont
  {Drozdov}}, \bibinfo {author} {\bibfnamefont {M.~I.}\ \bibnamefont
  {Eremets}}, \bibinfo {author} {\bibfnamefont {I.~A.}\ \bibnamefont {Troyan}},
  \bibinfo {author} {\bibfnamefont {V.}~\bibnamefont {Ksenofontov}}, \ and\
  \bibinfo {author} {\bibfnamefont {S.~I.}\ \bibnamefont {Shylin}},\ }\href
  {\doibase 10.1038/nature14964} {\bibfield  {journal} {\bibinfo  {journal}
  {Nature}\ }\textbf {\bibinfo {volume} {525}},\ \bibinfo {pages} {73}
  (\bibinfo {year} {2015})}\BibitemShut {NoStop}%
\bibitem [{\citenamefont {Einaga}\ \emph {et~al.}(2016)\citenamefont {Einaga},
  \citenamefont {Sakata}, \citenamefont {Ishikawa}, \citenamefont {Shimizu},
  \citenamefont {Eremets}, \citenamefont {Drozdov}, \citenamefont {Troyan},
  \citenamefont {Hirao},\ and\ \citenamefont {Ohishi}}]{einaga_crystal_2016}%
  \BibitemOpen
  \bibfield  {author} {\bibinfo {author} {\bibfnamefont {M.}~\bibnamefont
  {Einaga}}, \bibinfo {author} {\bibfnamefont {M.}~\bibnamefont {Sakata}},
  \bibinfo {author} {\bibfnamefont {T.}~\bibnamefont {Ishikawa}}, \bibinfo
  {author} {\bibfnamefont {K.}~\bibnamefont {Shimizu}}, \bibinfo {author}
  {\bibfnamefont {M.~I.}\ \bibnamefont {Eremets}}, \bibinfo {author}
  {\bibfnamefont {A.~P.}\ \bibnamefont {Drozdov}}, \bibinfo {author}
  {\bibfnamefont {I.~A.}\ \bibnamefont {Troyan}}, \bibinfo {author}
  {\bibfnamefont {N.}~\bibnamefont {Hirao}}, \ and\ \bibinfo {author}
  {\bibfnamefont {Y.}~\bibnamefont {Ohishi}},\ }\href {\doibase
  10.1038/nphys3760} {\bibfield  {journal} {\bibinfo  {journal} {Nature Phys}\
  }\textbf {\bibinfo {volume} {12}},\ \bibinfo {pages} {835} (\bibinfo {year}
  {2016})}\BibitemShut {NoStop}%
\bibitem [{\citenamefont {Ma}\ \emph {et~al.}(2022)\citenamefont {Ma},
  \citenamefont {Wang}, \citenamefont {Xie}, \citenamefont {Yang},
  \citenamefont {Wang}, \citenamefont {Zhou}, \citenamefont {Liu},
  \citenamefont {Yu}, \citenamefont {Zhao}, \citenamefont {Wang}, \citenamefont
  {Liu},\ and\ \citenamefont {Ma}}]{ma_high-tc_2021}%
  \BibitemOpen
  \bibfield  {author} {\bibinfo {author} {\bibfnamefont {L.}~\bibnamefont
  {Ma}}, \bibinfo {author} {\bibfnamefont {K.}~\bibnamefont {Wang}}, \bibinfo
  {author} {\bibfnamefont {Y.}~\bibnamefont {Xie}}, \bibinfo {author}
  {\bibfnamefont {X.}~\bibnamefont {Yang}}, \bibinfo {author} {\bibfnamefont
  {Y.}~\bibnamefont {Wang}}, \bibinfo {author} {\bibfnamefont {M.}~\bibnamefont
  {Zhou}}, \bibinfo {author} {\bibfnamefont {H.}~\bibnamefont {Liu}}, \bibinfo
  {author} {\bibfnamefont {X.}~\bibnamefont {Yu}}, \bibinfo {author}
  {\bibfnamefont {Y.}~\bibnamefont {Zhao}}, \bibinfo {author} {\bibfnamefont
  {H.}~\bibnamefont {Wang}}, \bibinfo {author} {\bibfnamefont {G.}~\bibnamefont
  {Liu}}, \ and\ \bibinfo {author} {\bibfnamefont {Y.}~\bibnamefont {Ma}},\
  }\href {\doibase 10.1103/PhysRevLett.128.167001} {\bibfield  {journal}
  {\bibinfo  {journal} {Phys. Rev. Lett.}\ }\textbf {\bibinfo {volume} {128}},\
  \bibinfo {pages} {167001} (\bibinfo {year} {2022})}\BibitemShut {NoStop}%
\bibitem [{\citenamefont {Chen}\ \emph {et~al.}(2021)\citenamefont {Chen},
  \citenamefont {Semenok}, \citenamefont {Huang}, \citenamefont {Shu},
  \citenamefont {Li}, \citenamefont {Duan}, \citenamefont {Cui},\ and\
  \citenamefont {Oganov}}]{chen_high-temperature_2021}%
  \BibitemOpen
  \bibfield  {author} {\bibinfo {author} {\bibfnamefont {W.}~\bibnamefont
  {Chen}}, \bibinfo {author} {\bibfnamefont {D.~V.}\ \bibnamefont {Semenok}},
  \bibinfo {author} {\bibfnamefont {X.}~\bibnamefont {Huang}}, \bibinfo
  {author} {\bibfnamefont {H.}~\bibnamefont {Shu}}, \bibinfo {author}
  {\bibfnamefont {X.}~\bibnamefont {Li}}, \bibinfo {author} {\bibfnamefont
  {D.}~\bibnamefont {Duan}}, \bibinfo {author} {\bibfnamefont {T.}~\bibnamefont
  {Cui}}, \ and\ \bibinfo {author} {\bibfnamefont {A.~R.}\ \bibnamefont
  {Oganov}},\ }\href {\doibase 10.1103/PhysRevLett.127.117001} {\bibfield
  {journal} {\bibinfo  {journal} {Phys. Rev. Lett.}\ }\textbf {\bibinfo
  {volume} {127}},\ \bibinfo {pages} {117001} (\bibinfo {year}
  {2021})}\BibitemShut {NoStop}%
\bibitem [{\citenamefont {Snider}\ \emph {et~al.}(2020)\citenamefont {Snider},
  \citenamefont {Dasenbrock-Gammon}, \citenamefont {McBride}, \citenamefont
  {Debessai}, \citenamefont {Vindana}, \citenamefont {Vencatasamy},
  \citenamefont {Lawler}, \citenamefont {Salamat},\ and\ \citenamefont
  {Dias}}]{snider_room-temperature_2020}%
  \BibitemOpen
  \bibfield  {author} {\bibinfo {author} {\bibfnamefont {E.}~\bibnamefont
  {Snider}}, \bibinfo {author} {\bibfnamefont {N.}~\bibnamefont
  {Dasenbrock-Gammon}}, \bibinfo {author} {\bibfnamefont {R.}~\bibnamefont
  {McBride}}, \bibinfo {author} {\bibfnamefont {M.}~\bibnamefont {Debessai}},
  \bibinfo {author} {\bibfnamefont {H.}~\bibnamefont {Vindana}}, \bibinfo
  {author} {\bibfnamefont {K.}~\bibnamefont {Vencatasamy}}, \bibinfo {author}
  {\bibfnamefont {K.~V.}\ \bibnamefont {Lawler}}, \bibinfo {author}
  {\bibfnamefont {A.}~\bibnamefont {Salamat}}, \ and\ \bibinfo {author}
  {\bibfnamefont {R.~P.}\ \bibnamefont {Dias}},\ }\href {\doibase
  10.1038/s41586-020-2801-z} {\bibfield  {journal} {\bibinfo  {journal}
  {Nature}\ }\textbf {\bibinfo {volume} {586}},\ \bibinfo {pages} {373}
  (\bibinfo {year} {2020})}\BibitemShut {NoStop}%
\bibitem [{\citenamefont {Semenok}\ \emph {et~al.}(2020)\citenamefont
  {Semenok}, \citenamefont {Kvashnin}, \citenamefont {Ivanova}, \citenamefont
  {Svitlyk}, \citenamefont {Fominski}, \citenamefont {Sadakov}, \citenamefont
  {Sobolevskiy}, \citenamefont {Pudalov}, \citenamefont {Troyan},\ and\
  \citenamefont {Oganov}}]{semenok_superconductivity_2020}%
  \BibitemOpen
  \bibfield  {author} {\bibinfo {author} {\bibfnamefont {D.~V.}\ \bibnamefont
  {Semenok}}, \bibinfo {author} {\bibfnamefont {A.~G.}\ \bibnamefont
  {Kvashnin}}, \bibinfo {author} {\bibfnamefont {A.~G.}\ \bibnamefont
  {Ivanova}}, \bibinfo {author} {\bibfnamefont {V.}~\bibnamefont {Svitlyk}},
  \bibinfo {author} {\bibfnamefont {V.~Y.}\ \bibnamefont {Fominski}}, \bibinfo
  {author} {\bibfnamefont {A.~V.}\ \bibnamefont {Sadakov}}, \bibinfo {author}
  {\bibfnamefont {O.~A.}\ \bibnamefont {Sobolevskiy}}, \bibinfo {author}
  {\bibfnamefont {V.~M.}\ \bibnamefont {Pudalov}}, \bibinfo {author}
  {\bibfnamefont {I.~A.}\ \bibnamefont {Troyan}}, \ and\ \bibinfo {author}
  {\bibfnamefont {A.~R.}\ \bibnamefont {Oganov}},\ }\href {\doibase
  10.1016/j.mattod.2019.10.005} {\bibfield  {journal} {\bibinfo  {journal}
  {Materials Today}\ }\textbf {\bibinfo {volume} {33}},\ \bibinfo {pages} {36}
  (\bibinfo {year} {2020})}\BibitemShut {NoStop}%
\bibitem [{\citenamefont {Semenok}\ \emph {et~al.}(2021)\citenamefont
  {Semenok}, \citenamefont {Troyan}, \citenamefont {Ivanova}, \citenamefont
  {Kvashnin}, \citenamefont {Kruglov}, \citenamefont {Hanfland}, \citenamefont
  {Sadakov}, \citenamefont {Sobolevskiy}, \citenamefont {Pervakov},
  \citenamefont {Lyubutin}, \citenamefont {Glazyrin}, \citenamefont {Giordano},
  \citenamefont {Karimov}, \citenamefont {Vasiliev}, \citenamefont {Akashi},
  \citenamefont {Pudalov},\ and\ \citenamefont
  {Oganov}}]{semenok_superconductivity_2021}%
  \BibitemOpen
  \bibfield  {author} {\bibinfo {author} {\bibfnamefont {D.~V.}\ \bibnamefont
  {Semenok}}, \bibinfo {author} {\bibfnamefont {I.~A.}\ \bibnamefont {Troyan}},
  \bibinfo {author} {\bibfnamefont {A.~G.}\ \bibnamefont {Ivanova}}, \bibinfo
  {author} {\bibfnamefont {A.~G.}\ \bibnamefont {Kvashnin}}, \bibinfo {author}
  {\bibfnamefont {I.~A.}\ \bibnamefont {Kruglov}}, \bibinfo {author}
  {\bibfnamefont {M.}~\bibnamefont {Hanfland}}, \bibinfo {author}
  {\bibfnamefont {A.~V.}\ \bibnamefont {Sadakov}}, \bibinfo {author}
  {\bibfnamefont {O.~A.}\ \bibnamefont {Sobolevskiy}}, \bibinfo {author}
  {\bibfnamefont {K.~S.}\ \bibnamefont {Pervakov}}, \bibinfo {author}
  {\bibfnamefont {I.~S.}\ \bibnamefont {Lyubutin}}, \bibinfo {author}
  {\bibfnamefont {K.~V.}\ \bibnamefont {Glazyrin}}, \bibinfo {author}
  {\bibfnamefont {N.}~\bibnamefont {Giordano}}, \bibinfo {author}
  {\bibfnamefont {D.~N.}\ \bibnamefont {Karimov}}, \bibinfo {author}
  {\bibfnamefont {A.~L.}\ \bibnamefont {Vasiliev}}, \bibinfo {author}
  {\bibfnamefont {R.}~\bibnamefont {Akashi}}, \bibinfo {author} {\bibfnamefont
  {V.~M.}\ \bibnamefont {Pudalov}}, \ and\ \bibinfo {author} {\bibfnamefont
  {A.~R.}\ \bibnamefont {Oganov}},\ }\href {\doibase
  10.1016/j.mattod.2021.03.025} {\bibfield  {journal} {\bibinfo  {journal}
  {Materials Today}\ ,\ \bibinfo {pages} {S1369702121001309}} (\bibinfo {year}
  {2021})}\BibitemShut {NoStop}%
\bibitem [{\citenamefont {Matsuoka}\ \emph {et~al.}(2019)\citenamefont
  {Matsuoka}, \citenamefont {Hishida}, \citenamefont {Kuno}, \citenamefont
  {Hirao}, \citenamefont {Ohishi}, \citenamefont {Sasaki}, \citenamefont
  {Takahama},\ and\ \citenamefont {Shimizu}}]{matsuoka_superconductivity_2019}%
  \BibitemOpen
  \bibfield  {author} {\bibinfo {author} {\bibfnamefont {T.}~\bibnamefont
  {Matsuoka}}, \bibinfo {author} {\bibfnamefont {M.}~\bibnamefont {Hishida}},
  \bibinfo {author} {\bibfnamefont {K.}~\bibnamefont {Kuno}}, \bibinfo {author}
  {\bibfnamefont {N.}~\bibnamefont {Hirao}}, \bibinfo {author} {\bibfnamefont
  {Y.}~\bibnamefont {Ohishi}}, \bibinfo {author} {\bibfnamefont
  {S.}~\bibnamefont {Sasaki}}, \bibinfo {author} {\bibfnamefont
  {K.}~\bibnamefont {Takahama}}, \ and\ \bibinfo {author} {\bibfnamefont
  {K.}~\bibnamefont {Shimizu}},\ }\href {\doibase 10.1103/PhysRevB.99.144511}
  {\bibfield  {journal} {\bibinfo  {journal} {Phys. Rev. B}\ }\textbf {\bibinfo
  {volume} {99}},\ \bibinfo {pages} {144511} (\bibinfo {year}
  {2019})}\BibitemShut {NoStop}%
\bibitem [{\citenamefont {Somayazulu}\ \emph {et~al.}(2019)\citenamefont
  {Somayazulu}, \citenamefont {Ahart}, \citenamefont {Mishra}, \citenamefont
  {Geballe}, \citenamefont {Baldini}, \citenamefont {Meng}, \citenamefont
  {Struzhkin},\ and\ \citenamefont {Hemley}}]{somayazulu_evidence_2019}%
  \BibitemOpen
  \bibfield  {author} {\bibinfo {author} {\bibfnamefont {M.}~\bibnamefont
  {Somayazulu}}, \bibinfo {author} {\bibfnamefont {M.}~\bibnamefont {Ahart}},
  \bibinfo {author} {\bibfnamefont {A.~K.}\ \bibnamefont {Mishra}}, \bibinfo
  {author} {\bibfnamefont {Z.~M.}\ \bibnamefont {Geballe}}, \bibinfo {author}
  {\bibfnamefont {M.}~\bibnamefont {Baldini}}, \bibinfo {author} {\bibfnamefont
  {Y.}~\bibnamefont {Meng}}, \bibinfo {author} {\bibfnamefont {V.~V.}\
  \bibnamefont {Struzhkin}}, \ and\ \bibinfo {author} {\bibfnamefont {R.~J.}\
  \bibnamefont {Hemley}},\ }\href {\doibase 10.1103/PhysRevLett.122.027001}
  {\bibfield  {journal} {\bibinfo  {journal} {Phys. Rev. Lett.}\ }\textbf
  {\bibinfo {volume} {122}},\ \bibinfo {pages} {027001} (\bibinfo {year}
  {2019})}\BibitemShut {NoStop}%
\bibitem [{\citenamefont {Drozdov}\ \emph {et~al.}(2019)\citenamefont
  {Drozdov}, \citenamefont {Kong}, \citenamefont {Minkov}, \citenamefont
  {Besedin}, \citenamefont {Kuzovnikov}, \citenamefont {Mozaffari},
  \citenamefont {Balicas}, \citenamefont {Balakirev}, \citenamefont {Graf},
  \citenamefont {Prakapenka}, \citenamefont {Greenberg}, \citenamefont
  {Knyazev}, \citenamefont {Tkacz},\ and\ \citenamefont
  {Eremets}}]{drozdov_superconductivity_2019}%
  \BibitemOpen
  \bibfield  {author} {\bibinfo {author} {\bibfnamefont {A.~P.}\ \bibnamefont
  {Drozdov}}, \bibinfo {author} {\bibfnamefont {P.~P.}\ \bibnamefont {Kong}},
  \bibinfo {author} {\bibfnamefont {V.~S.}\ \bibnamefont {Minkov}}, \bibinfo
  {author} {\bibfnamefont {S.~P.}\ \bibnamefont {Besedin}}, \bibinfo {author}
  {\bibfnamefont {M.~A.}\ \bibnamefont {Kuzovnikov}}, \bibinfo {author}
  {\bibfnamefont {S.}~\bibnamefont {Mozaffari}}, \bibinfo {author}
  {\bibfnamefont {L.}~\bibnamefont {Balicas}}, \bibinfo {author} {\bibfnamefont
  {F.~F.}\ \bibnamefont {Balakirev}}, \bibinfo {author} {\bibfnamefont {D.~E.}\
  \bibnamefont {Graf}}, \bibinfo {author} {\bibfnamefont {V.~B.}\ \bibnamefont
  {Prakapenka}}, \bibinfo {author} {\bibfnamefont {E.}~\bibnamefont
  {Greenberg}}, \bibinfo {author} {\bibfnamefont {D.~A.}\ \bibnamefont
  {Knyazev}}, \bibinfo {author} {\bibfnamefont {M.}~\bibnamefont {Tkacz}}, \
  and\ \bibinfo {author} {\bibfnamefont {M.~I.}\ \bibnamefont {Eremets}},\
  }\href {\doibase 10.1038/s41586-019-1201-8} {\bibfield  {journal} {\bibinfo
  {journal} {Nature}\ }\textbf {\bibinfo {volume} {569}},\ \bibinfo {pages}
  {528} (\bibinfo {year} {2019})}\BibitemShut {NoStop}%
\bibitem [{\citenamefont {Sun}\ \emph {et~al.}(2021)\citenamefont {Sun},
  \citenamefont {Minkov}, \citenamefont {Mozaffari}, \citenamefont {Sun},
  \citenamefont {Ma}, \citenamefont {Chariton}, \citenamefont {Prakapenka},
  \citenamefont {Eremets}, \citenamefont {Balicas},\ and\ \citenamefont
  {Balakirev}}]{sun_high-temperature_2021}%
  \BibitemOpen
  \bibfield  {author} {\bibinfo {author} {\bibfnamefont {D.}~\bibnamefont
  {Sun}}, \bibinfo {author} {\bibfnamefont {V.~S.}\ \bibnamefont {Minkov}},
  \bibinfo {author} {\bibfnamefont {S.}~\bibnamefont {Mozaffari}}, \bibinfo
  {author} {\bibfnamefont {Y.}~\bibnamefont {Sun}}, \bibinfo {author}
  {\bibfnamefont {Y.}~\bibnamefont {Ma}}, \bibinfo {author} {\bibfnamefont
  {S.}~\bibnamefont {Chariton}}, \bibinfo {author} {\bibfnamefont {V.~B.}\
  \bibnamefont {Prakapenka}}, \bibinfo {author} {\bibfnamefont {M.~I.}\
  \bibnamefont {Eremets}}, \bibinfo {author} {\bibfnamefont {L.}~\bibnamefont
  {Balicas}}, \ and\ \bibinfo {author} {\bibfnamefont {F.~F.}\ \bibnamefont
  {Balakirev}},\ }\href {\doibase 10.1038/s41467-021-26706-w} {\bibfield
  {journal} {\bibinfo  {journal} {Nature Communications}\ }\textbf {\bibinfo
  {volume} {12}},\ \bibinfo {pages} {6863} (\bibinfo {year}
  {2021})}\BibitemShut {NoStop}%
\bibitem [{\citenamefont {Geballe}\ \emph {et~al.}(2018)\citenamefont
  {Geballe}, \citenamefont {Liu}, \citenamefont {Mishra}, \citenamefont
  {Ahart}, \citenamefont {Somayazulu}, \citenamefont {Meng}, \citenamefont
  {Baldini},\ and\ \citenamefont {Hemley}}]{geballe_synthesis_2018}%
  \BibitemOpen
  \bibfield  {author} {\bibinfo {author} {\bibfnamefont {Z.~M.}\ \bibnamefont
  {Geballe}}, \bibinfo {author} {\bibfnamefont {H.}~\bibnamefont {Liu}},
  \bibinfo {author} {\bibfnamefont {A.~K.}\ \bibnamefont {Mishra}}, \bibinfo
  {author} {\bibfnamefont {M.}~\bibnamefont {Ahart}}, \bibinfo {author}
  {\bibfnamefont {M.}~\bibnamefont {Somayazulu}}, \bibinfo {author}
  {\bibfnamefont {Y.}~\bibnamefont {Meng}}, \bibinfo {author} {\bibfnamefont
  {M.}~\bibnamefont {Baldini}}, \ and\ \bibinfo {author} {\bibfnamefont
  {R.~J.}\ \bibnamefont {Hemley}},\ }\href {\doibase 10.1002/anie.201709970}
  {\bibfield  {journal} {\bibinfo  {journal} {Angew. Chem. Int. Ed.}\ }\textbf
  {\bibinfo {volume} {57}},\ \bibinfo {pages} {688} (\bibinfo {year}
  {2018})}\BibitemShut {NoStop}%
\bibitem [{\citenamefont {Errea}\ \emph {et~al.}(2020)\citenamefont {Errea},
  \citenamefont {Belli}, \citenamefont {Monacelli}, \citenamefont {Sanna},
  \citenamefont {Koretsune}, \citenamefont {Tadano}, \citenamefont {Bianco},
  \citenamefont {Calandra}, \citenamefont {Arita}, \citenamefont {Mauri},\ and\
  \citenamefont {Flores-Livas}}]{errea_quantum_2020}%
  \BibitemOpen
  \bibfield  {author} {\bibinfo {author} {\bibfnamefont {I.}~\bibnamefont
  {Errea}}, \bibinfo {author} {\bibfnamefont {F.}~\bibnamefont {Belli}},
  \bibinfo {author} {\bibfnamefont {L.}~\bibnamefont {Monacelli}}, \bibinfo
  {author} {\bibfnamefont {A.}~\bibnamefont {Sanna}}, \bibinfo {author}
  {\bibfnamefont {T.}~\bibnamefont {Koretsune}}, \bibinfo {author}
  {\bibfnamefont {T.}~\bibnamefont {Tadano}}, \bibinfo {author} {\bibfnamefont
  {R.}~\bibnamefont {Bianco}}, \bibinfo {author} {\bibfnamefont
  {M.}~\bibnamefont {Calandra}}, \bibinfo {author} {\bibfnamefont
  {R.}~\bibnamefont {Arita}}, \bibinfo {author} {\bibfnamefont
  {F.}~\bibnamefont {Mauri}}, \ and\ \bibinfo {author} {\bibfnamefont {J.~A.}\
  \bibnamefont {Flores-Livas}},\ }\href {\doibase 10.1038/s41586-020-1955-z}
  {\bibfield  {journal} {\bibinfo  {journal} {Nature}\ }\textbf {\bibinfo
  {volume} {578}},\ \bibinfo {pages} {66} (\bibinfo {year} {2020})}\BibitemShut
  {NoStop}%
\bibitem [{\citenamefont {Wang}\ \emph {et~al.}(2021)\citenamefont {Wang},
  \citenamefont {Yao}, \citenamefont {Peng}, \citenamefont {Liu},\ and\
  \citenamefont {Hemley}}]{wang_quantum_2021}%
  \BibitemOpen
  \bibfield  {author} {\bibinfo {author} {\bibfnamefont {H.}~\bibnamefont
  {Wang}}, \bibinfo {author} {\bibfnamefont {Y.}~\bibnamefont {Yao}}, \bibinfo
  {author} {\bibfnamefont {F.}~\bibnamefont {Peng}}, \bibinfo {author}
  {\bibfnamefont {H.}~\bibnamefont {Liu}}, \ and\ \bibinfo {author}
  {\bibfnamefont {R.}~\bibnamefont {Hemley}},\ }\href {\doibase
  10.1103/PhysRevLett.126.117002} {\bibfield  {journal} {\bibinfo  {journal}
  {Phys. Rev. Lett.}\ }\textbf {\bibinfo {volume} {126}},\ \bibinfo {pages}
  {117002} (\bibinfo {year} {2021})}\BibitemShut {NoStop}%
\bibitem [{\citenamefont {Liu}\ \emph {et~al.}(2018)\citenamefont {Liu},
  \citenamefont {Naumov}, \citenamefont {Geballe}, \citenamefont {Somayazulu},
  \citenamefont {Tse},\ and\ \citenamefont {Hemley}}]{liu_dynamics_2018}%
  \BibitemOpen
  \bibfield  {author} {\bibinfo {author} {\bibfnamefont {H.}~\bibnamefont
  {Liu}}, \bibinfo {author} {\bibfnamefont {I.~I.}\ \bibnamefont {Naumov}},
  \bibinfo {author} {\bibfnamefont {Z.~M.}\ \bibnamefont {Geballe}}, \bibinfo
  {author} {\bibfnamefont {M.}~\bibnamefont {Somayazulu}}, \bibinfo {author}
  {\bibfnamefont {J.~S.}\ \bibnamefont {Tse}}, \ and\ \bibinfo {author}
  {\bibfnamefont {R.~J.}\ \bibnamefont {Hemley}},\ }\href {\doibase
  10.1103/PhysRevB.98.100102} {\bibfield  {journal} {\bibinfo  {journal} {Phys.
  Rev. B}\ }\textbf {\bibinfo {volume} {98}},\ \bibinfo {pages} {100102}
  (\bibinfo {year} {2018})}\BibitemShut {NoStop}%
\bibitem [{Note1()}]{Note1}%
  \BibitemOpen
  \bibinfo {note} {For MD and PIMD simulations to discuss the dynamical
  properties of $\protect \mathrm {LaH_{10}}$, see Ref.~\protect
  \rev@citealpnum {liu_dynamics_2018}}\BibitemShut {NoStop}%
\bibitem [{\citenamefont {Shiga}\ \emph {et~al.}(2001)\citenamefont {Shiga},
  \citenamefont {Tachikawa},\ and\ \citenamefont {Miura}}]{shiga_unified_2001}%
  \BibitemOpen
  \bibfield  {author} {\bibinfo {author} {\bibfnamefont {M.}~\bibnamefont
  {Shiga}}, \bibinfo {author} {\bibfnamefont {M.}~\bibnamefont {Tachikawa}}, \
  and\ \bibinfo {author} {\bibfnamefont {S.}~\bibnamefont {Miura}},\ }\href
  {\doibase 10.1063/1.1407289} {\bibfield  {journal} {\bibinfo  {journal} {The
  Journal of Chemical Physics}\ }\textbf {\bibinfo {volume} {115}},\ \bibinfo
  {pages} {9149} (\bibinfo {year} {2001})}\BibitemShut {NoStop}%
\bibitem [{\citenamefont {Tuckerman}\ \emph {et~al.}(1993)\citenamefont
  {Tuckerman}, \citenamefont {Berne}, \citenamefont {Martyna},\ and\
  \citenamefont {Klein}}]{tuckerman}%
  \BibitemOpen
  \bibfield  {author} {\bibinfo {author} {\bibfnamefont {M.~E.}\ \bibnamefont
  {Tuckerman}}, \bibinfo {author} {\bibfnamefont {B.~J.}\ \bibnamefont
  {Berne}}, \bibinfo {author} {\bibfnamefont {G.~J.}\ \bibnamefont {Martyna}},
  \ and\ \bibinfo {author} {\bibfnamefont {M.~L.}\ \bibnamefont {Klein}},\
  }\href {\doibase 10.1063/1.465188} {\bibfield  {journal} {\bibinfo  {journal}
  {The Journal of Chemical Physics}\ }\textbf {\bibinfo {volume} {99}},\
  \bibinfo {pages} {2796} (\bibinfo {year} {1993})}\BibitemShut {NoStop}%
\bibitem [{\citenamefont {Marx}\ and\ \citenamefont {Parrinello}(1996)}]{marx}%
  \BibitemOpen
  \bibfield  {author} {\bibinfo {author} {\bibfnamefont {D.}~\bibnamefont
  {Marx}}\ and\ \bibinfo {author} {\bibfnamefont {M.}~\bibnamefont
  {Parrinello}},\ }\href {\doibase 10.1063/1.471221} {\bibfield  {journal}
  {\bibinfo  {journal} {The Journal of Chemical Physics}\ }\textbf {\bibinfo
  {volume} {104}},\ \bibinfo {pages} {4077} (\bibinfo {year}
  {1996})}\BibitemShut {NoStop}%
\bibitem [{\citenamefont {Shiga}(2020)}]{pimd}%
  \BibitemOpen
  \bibfield  {author} {\bibinfo {author} {\bibfnamefont {M.}~\bibnamefont
  {Shiga}},\ }\href@noop {} {\bibfield  {journal} {\bibinfo  {journal} {PIMD}\
  } (\bibinfo {year} {2020})}\BibitemShut {NoStop}%
\bibitem [{\citenamefont {Kresse}\ and\ \citenamefont
  {Furthmüller}(1996)}]{kresse_efficient_1996}%
  \BibitemOpen
  \bibfield  {author} {\bibinfo {author} {\bibfnamefont {G.}~\bibnamefont
  {Kresse}}\ and\ \bibinfo {author} {\bibfnamefont {J.}~\bibnamefont
  {Furthmüller}},\ }\href {\doibase 10.1103/PhysRevB.54.11169} {\bibfield
  {journal} {\bibinfo  {journal} {Phys. Rev. B}\ }\textbf {\bibinfo {volume}
  {54}},\ \bibinfo {pages} {11169} (\bibinfo {year} {1996})}\BibitemShut
  {NoStop}%
\bibitem [{\citenamefont {Kresse}\ and\ \citenamefont
  {Joubert}(1999)}]{kresse_ultrasoft_1999}%
  \BibitemOpen
  \bibfield  {author} {\bibinfo {author} {\bibfnamefont {G.}~\bibnamefont
  {Kresse}}\ and\ \bibinfo {author} {\bibfnamefont {D.}~\bibnamefont
  {Joubert}},\ }\href {\doibase 10.1103/PhysRevB.59.1758} {\bibfield  {journal}
  {\bibinfo  {journal} {Phys. Rev. B}\ }\textbf {\bibinfo {volume} {59}},\
  \bibinfo {pages} {1758} (\bibinfo {year} {1999})}\BibitemShut {NoStop}%
\bibitem [{\citenamefont {Bl{\"o}chl}(1994)}]{blochl_projector_1994}%
  \BibitemOpen
  \bibfield  {author} {\bibinfo {author} {\bibfnamefont {P.~E.}\ \bibnamefont
  {Bl{\"o}chl}},\ }\href {\doibase 10.1103/PhysRevB.50.17953} {\bibfield
  {journal} {\bibinfo  {journal} {Phys. Rev. B}\ }\textbf {\bibinfo {volume}
  {50}},\ \bibinfo {pages} {17953} (\bibinfo {year} {1994})}\BibitemShut
  {NoStop}%
\bibitem [{\citenamefont {Perdew}\ \emph {et~al.}(1996)\citenamefont {Perdew},
  \citenamefont {Burke},\ and\ \citenamefont
  {Ernzerhof}}]{perdew_generalized_1996}%
  \BibitemOpen
  \bibfield  {author} {\bibinfo {author} {\bibfnamefont {J.~P.}\ \bibnamefont
  {Perdew}}, \bibinfo {author} {\bibfnamefont {K.}~\bibnamefont {Burke}}, \
  and\ \bibinfo {author} {\bibfnamefont {M.}~\bibnamefont {Ernzerhof}},\ }\href
  {\doibase 10.1103/PhysRevLett.77.3865} {\bibfield  {journal} {\bibinfo
  {journal} {Phys. Rev. Lett.}\ }\textbf {\bibinfo {volume} {77}},\ \bibinfo
  {pages} {3865} (\bibinfo {year} {1996})}\BibitemShut {NoStop}%
\bibitem [{Note2()}]{Note2}%
  \BibitemOpen
  \bibinfo {note} {For the dependence of the critical pressure at the
  structural phase transition on the choice of the pseudopotentials and
  exchange-correlation functionals, see Ref.~\protect \rev@citealpnum
  {sura_effects_2022}}\BibitemShut {NoStop}%
\bibitem [{\citenamefont {Lin}\ \emph {et~al.}(2003)\citenamefont {Lin},
  \citenamefont {Blanco},\ and\ \citenamefont {Goddard}}]{entropy}%
  \BibitemOpen
  \bibfield  {author} {\bibinfo {author} {\bibfnamefont {S.-T.}\ \bibnamefont
  {Lin}}, \bibinfo {author} {\bibfnamefont {M.}~\bibnamefont {Blanco}}, \ and\
  \bibinfo {author} {\bibfnamefont {W.~A.}\ \bibnamefont {Goddard}},\ }\href
  {\doibase 10.1063/1.1624057} {\bibfield  {journal} {\bibinfo  {journal} {The
  Journal of Chemical Physics}\ }\textbf {\bibinfo {volume} {119}},\ \bibinfo
  {pages} {11792} (\bibinfo {year} {2003})},\ \Eprint
  {http://arxiv.org/abs/https://doi.org/10.1063/1.1624057}
  {https://doi.org/10.1063/1.1624057} \BibitemShut {NoStop}%
\bibitem [{\citenamefont {Momma}\ and\ \citenamefont {Izumi}(2011)}]{vesta}%
  \BibitemOpen
  \bibfield  {author} {\bibinfo {author} {\bibfnamefont {K.}~\bibnamefont
  {Momma}}\ and\ \bibinfo {author} {\bibfnamefont {F.}~\bibnamefont {Izumi}},\
  }\href@noop {} {\bibfield  {journal} {\bibinfo  {journal} {J. Appl.
  Crystallogr.}\ }\textbf {\bibinfo {volume} {44}},\ \bibinfo {pages} {1272}
  (\bibinfo {year} {2011})}\BibitemShut {NoStop}%
\bibitem [{\citenamefont {Ong}\ \emph {et~al.}(2013)\citenamefont {Ong},
  \citenamefont {Richards}, \citenamefont {Jain}, \citenamefont {Hautier},
  \citenamefont {Kocher}, \citenamefont {Cholia}, \citenamefont {Gunter},
  \citenamefont {Chevrier}, \citenamefont {Persson},\ and\ \citenamefont
  {Ceder}}]{ong_python_2013}%
  \BibitemOpen
  \bibfield  {author} {\bibinfo {author} {\bibfnamefont {S.~P.}\ \bibnamefont
  {Ong}}, \bibinfo {author} {\bibfnamefont {W.~D.}\ \bibnamefont {Richards}},
  \bibinfo {author} {\bibfnamefont {A.}~\bibnamefont {Jain}}, \bibinfo {author}
  {\bibfnamefont {G.}~\bibnamefont {Hautier}}, \bibinfo {author} {\bibfnamefont
  {M.}~\bibnamefont {Kocher}}, \bibinfo {author} {\bibfnamefont
  {S.}~\bibnamefont {Cholia}}, \bibinfo {author} {\bibfnamefont
  {D.}~\bibnamefont {Gunter}}, \bibinfo {author} {\bibfnamefont {V.~L.}\
  \bibnamefont {Chevrier}}, \bibinfo {author} {\bibfnamefont {K.~A.}\
  \bibnamefont {Persson}}, \ and\ \bibinfo {author} {\bibfnamefont
  {G.}~\bibnamefont {Ceder}},\ }\href {\doibase
  10.1016/j.commatsci.2012.10.028} {\bibfield  {journal} {\bibinfo  {journal}
  {Computational Materials Science}\ }\textbf {\bibinfo {volume} {68}},\
  \bibinfo {pages} {314} (\bibinfo {year} {2013})}\BibitemShut {NoStop}%
\bibitem [{\citenamefont {Togo}\ and\ \citenamefont
  {Tanaka}(2018)}]{togo_textttspglib_2018}%
  \BibitemOpen
  \bibfield  {author} {\bibinfo {author} {\bibfnamefont {A.}~\bibnamefont
  {Togo}}\ and\ \bibinfo {author} {\bibfnamefont {I.}~\bibnamefont {Tanaka}},\
  }\href@noop {} {\bibfield  {journal} {\bibinfo  {journal} {arXiv:1808.01590
  [cond-mat]}\ } (\bibinfo {year} {2018})},\ \bibinfo {note} {arXiv:
  1808.01590}\BibitemShut {NoStop}%
\bibitem [{\citenamefont {Sura}\ \emph {et~al.}(2022)\citenamefont {Sura},
  \citenamefont {Verma},\ and\ \citenamefont {Garg}}]{sura_effects_2022}%
  \BibitemOpen
  \bibfield  {author} {\bibinfo {author} {\bibfnamefont {S.}~\bibnamefont
  {Sura}}, \bibinfo {author} {\bibfnamefont {A.~K.}\ \bibnamefont {Verma}}, \
  and\ \bibinfo {author} {\bibfnamefont {N.}~\bibnamefont {Garg}},\ }\href
  {\doibase 10.1016/j.ssc.2021.114583} {\bibfield  {journal} {\bibinfo
  {journal} {Solid State Communications}\ }\textbf {\bibinfo {volume} {341}},\
  \bibinfo {pages} {114583} (\bibinfo {year} {2022})}\BibitemShut {NoStop}%
\end{thebibliography}%

\end{document}